\documentclass[floatfix,aps,pra,superscriptaddress,longbibliography,twocolumn]{revtex4-2}
\usepackage{color}
\usepackage{xcolor}
\usepackage{amsmath}
\usepackage{amsthm}
\usepackage{amssymb}
\usepackage{graphicx}
\usepackage{makecell}
\usepackage{enumitem}
\usepackage{booktabs}
\usepackage{url}
\usepackage{hyperref}
\usepackage{lineno}

\graphicspath{{../MBE_version/pictures_final/}}

\begin{document}

\title{Reconstructability of evolutionary intermediates in generative epistatic landscapes}

\author{Roberto Netti}
\affiliation{Department of Computational, Quantitative and Synthetic Biology,
Sorbonne Universit\'e, CNRS, 75005 Paris, France}

\author{Martin Weigt}
\email{martin.weigt@gmail.com}
\affiliation{Department of Computational, Quantitative and Synthetic Biology,
Sorbonne Universit\'e, CNRS, 75005 Paris, France}
\affiliation{Institut Universitaire de France (IUF), France}

\begin{abstract}
Evolutionary intermediates connect observed proteins, but the sequence of steps that produced them is rarely recoverable from extant data alone. Here we ask what can, and cannot, be inferred about such intermediates from the endpoints. Using generative sequence landscapes as controlled models of protein-family evolution, we benchmark data-driven reconstruction against ground-truth simulated trajectories. We find that the best point prediction is not necessarily the most faithful evolutionary reconstruction: maximum-likelihood intermediates can be residue-wise accurate yet statistically atypical, whereas conditional sampling better captures the ensemble of plausible histories. Predictability is limited by the topology of the landscape. Constrained, low-mutability regions preserve information about the path, while permissive high-mutability regions open many alternative routes and erase path-specific memory. We also show that sequence divergence alone is an insufficient measure of elapsed evolutionary time; incorporating endpoint mutability provides a more reliable way to place intermediates in the landscape. These results recast intermediate reconstruction as a calibrated probabilistic problem. Rather than seeking a single ``true'' sequence, data-driven models should identify when endpoints contain evolutionary information, and return realistic ensembles.
\end{abstract}

\maketitle

\section{Introduction}

The computational design of proteins with tailored functionalities has witnessed a paradigm shift in recent years. Driven by the explosion of genomic data, generative probabilistic models, ranging from classical ``energy based models'' (like Potts models and restricted Boltzmann machines) \cite{Russ2020, mcgee2021generative, Rehan2026, Netti2026HighEntropy} to deep learning architectures (like variational autoencoders, graph neural networks and large language models) \cite{HawkinsHooker2021, Madani2023, ProteinMPNN2022, nijkamp2023progen2, Ferruz2022ProtGPT2}, have proven their efficiency when learning the statistical constraints of protein families and generating novel sequences, which differ substantially from any observed natural sequence but can be proven functional in experiments. However, while the problem of designing individual sequences using generative models as proxies for fitness landscapes is rather well-addressed, the problem of \textit{how proteins evolve in the sequence landscape} remains a significant open challenge. Specifically, identifying plausible evolutionary paths that connect two distinct wild-type sequences is of evident biological and bioengineering interest; however, it has been treated to a much lesser extent.

The main difficulty is that exhaustive experimental validation of evolutionary paths is generally infeasible, in contrast to static sequence designs. Between two wild-type sequences, the number of possible mutational paths grows rapidly with sequence divergence, while epistasis strongly restricts which paths are viable \cite{Poelwijk2019Epistasis, Weinreich2006}. This makes the choice and validation of plausible trajectories intrinsically difficult.

Several studies have attempted to generate or analyze paths between wild-type sequences with different functions, specificities, or folds, either by using model-derived fitness landscapes or by exploring the local combinatorial space between closely related sequences \cite{Rehan2026, Mauri2023, Kantroo2024, Voordeckers2012}. These works show that some intermediates can be functional or promiscuous, but also suggest that transitions between endpoints may involve stability trade-offs \cite{TianBest2020}.

Here, we take a complementary data-driven approach: instead of selecting paths directly, we learn intermediate reconstruction from simulated evolutionary trajectories generated within inferred family-specific landscapes \cite{DiBari2024}, cf.~also \cite{barton2016relative,Bisardi2022,biswas2024kinetic,alvarez2024vivo,Rossi2025} for related work on simulating protein evolution in data-driven sequence landscapes.

\begin{figure*}[!t]
    \centering
    \includegraphics[width=0.95\textwidth]{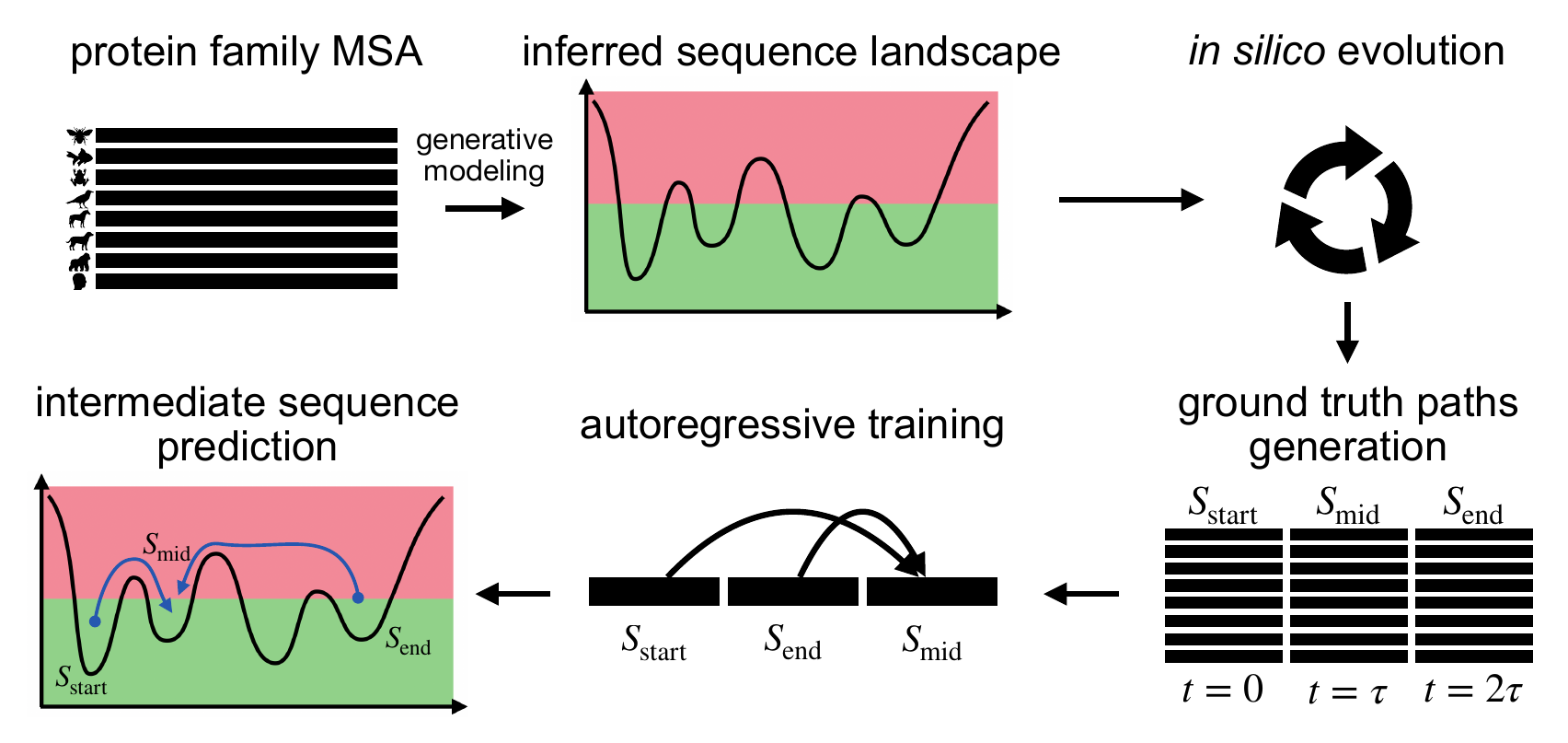}
    \caption{\textbf{Data-driven framework for reconstructing evolutionary intermediates.}
    \textit{(A) Landscape inference:} a generative probabilistic model is trained on natural multiple sequence alignments as a proxy for the epistatic fitness landscape of a protein family.
    \textit{(B) Simulated evolution:} mutational trajectories are simulated in this landscape, generating a ground-truth dataset of sequence triples, comprising a starting point $S_{\rm start}$, an intermediate midpoint $S_{\rm mid}$, and a final endpoint $S_{\rm end}$, across various evolutionary timescales.
    \textit{(C) Autoregressive training:} an autoregressive generative model is trained on these triples to learn the conditional probability $P(S_{\rm mid}\mid S_{\rm start},S_{\rm end})$ of the intermediate sequence given the initial and final ones.
    \textit{(D) Path prediction:} the trained model is deployed to generate statistically realistic intermediate sequences bridging distant wild-type proteins, using posterior sampling from $P(S_{\rm mid}\mid S_{\rm start},S_{\rm end})$.}
    \label{fig:graphical_abstract}
\end{figure*}

In particular, we aim at establishing a rigorous framework for dissecting the reconstructability of evolutionary paths. There are a few important obstacles, which have to be addressed:
\begin{itemize}
    \item \textit{The lack of ground truth data} is a recurrent problem in evolutionary studies. Databases are populated by extant sequences, but only few evolutionary trajectories or actual ancestral sequences are known \cite{Thornton2004, Hofreiter2001}. Exceptions are trajectories like in the case of evolving viral pathogens or for evolution experiments, but they are typically restricted to short evolutionary time scales and limited sequence divergence \cite{RomeroArnold2009, Fantini2020, Stiffler2020, Markov2023}.
    \item \textit{The lack of realism of simulated evolutionary trajectories}, which are frequently based on simplifying assumptions like independent-site evolution or site-homogeneous substitution rates, ignoring heterogeneous conservation patterns and the pervasive nature of epistasis in protein evolution. Overly simplified evolution models therefore lead, for long simulations, with high probability to non-viable protein sequences, which are statistically easily distinguishable from natural sequences \cite{DiBari2026, Weinreich2005, Starr2016, Buda2023}.
    \item \textit{The intrinsic stochasticity of natural evolution} based on random mutations makes a faithful reconstruction of individual deeply divergent sequences impossible, and potential ensembles of plausible sequences should replace deterministic reconstructions \cite{Park2022, Rossi2025}.
\end{itemize}

Note that many of these problems are shared with the related problem of ancestral sequence reconstruction \cite{yang1995new,pupko2000fast,HansonSmith2010,joy2016ancestral}. Here, we use the framework of generative probabilistic models mentioned above to address these issues, cf.~Fig.~\ref{fig:graphical_abstract}. Trained on the available extant data (Fig.~\ref{fig:graphical_abstract}A), these models are already used to address evolutionary questions like the quantification of selective effects of mutations and to enable evolutionary simulations matching experimental data statistics across divergence scales \cite{DiBari2024, Figliuzzi2016, Bisardi2022}. These advances allow us to address the first two issues and build our work on more realistic simulated trajectories (Fig.~\ref{fig:graphical_abstract}B). Based on massive simulated trajectories, we can train novel generative models for evolutionary intermediate sequences, simultaneously conditioned on more ancestral and more recent WT sequences (Fig.~\ref{fig:graphical_abstract}C). This approach allows us to generate ``natural-like'' evolutionary trajectories and systematically evaluate how the properties of the endpoint WTs $S_\mathrm{start}$ and $S_\mathrm{end}$ influence the predictability of the intermediate sequences called $S_\mathrm{mid}$.

Our study focuses on three primary objectives:
\begin{enumerate}
    \item We benchmark data-driven approaches on simulated data to assess the \textit{potential and limits} of different reconstruction strategies. We show that a naive direct-path baseline fails at longer timescales by ignoring epistasis and sites with multiple substitutions. We also show that while a deterministic maximum-likelihood reconstruction of an autoregressive model conditioned on $S_\mathrm{start}$ and $S_\mathrm{end}$ minimizes the reconstruction error of the evolutionary intermediate $S_\mathrm{mid}$, it does so at the cost of producing atypical ``low-cost'' paths. Conditional sampling from the learned conditional distribution remains essential to recover the typical behavior of evolutionary dynamics.
    \item We investigate the intrinsic limits of intermediate prediction imposed by the topology of the fitness landscape. We show that path information is not retained uniformly across the trajectories: low energy (high proxy fitness) regions remain more constrained and therefore preserve stronger memory of the evolutionary route, whereas high energy (low proxy fitness) regions allow many alternative mutational paths. When trajectories pass through such permissive regions, the memory of the realized path is progressively weakened, limiting the predictability of $S_\mathrm{mid}$ from $S_\mathrm{start}$ and $S_\mathrm{end}$ alone. These regions therefore act as an \textit{information horizon}, beyond which endpoint sequences no longer contain enough information to reliably reconstruct the specific intermediate.
    \item Finally, we suggest how to bridge the gap between \textit{artificial simulation and real-case application}. We demonstrate that while simple sequence divergence (Hamming distance of aligned sequences) is an insufficient proxy for evolutionary time in heterogeneous epistatic landscapes, it is possible to accurately infer the latent evolutionary timescale by integrating model-predicted mutational robustnesses of $S_\mathrm{start}$ and $S_\mathrm{end}$. This provides a practical bridge toward real protein data, offering a robust method for navigating the sequence landscape even when the real evolutionary history is unknown.
\end{enumerate}

We formulate our approach within the generative framework of Direct Coupling Analysis (DCA) \cite{Morcos2011, Cocco2018}, which has been demonstrated to allow the building of accurate generative models for individual families of homologous proteins and the generation of functional artificial biological sequences \cite{Russ2020, Netti2026HighEntropy, Calvanese2025}. The general ideas are, however, directly applicable to other generative modeling frameworks, including deep-learning architectures, in particular when formulated in an autoregressive way (which we use naturally for conditioning a model for $S_\mathrm{mid}$ to $S_\mathrm{start}$ and $S_\mathrm{end}$) \cite{Trinquier2021,Madani2023,nijkamp2023progen2,Ferruz2022ProtGPT2}. Our findings are illustrated in the Results section using one protein family, the chorismate mutases, as the main example. Further details on the datasets and on the proposed modeling framework are provided in \textit{Materials and Methods}. Additional results for the $\beta$-lactamase and RR domain families are reported in the \textit{Supplementary Appendix}, Section S6.

\section{Results}

\subsection{Benchmarking data-driven prediction of evolutionary intermediates on simulated trajectories}

To assess the potential and the limits of reconstructing evolutionary intermediates and evolutionary paths, we rely on controlled \textit{in silico} evolutionary data, following the quantitative method proposed in \cite{DiBari2024}. To do so,
\begin{enumerate}
    \item[$(i)$] we collect deep multiple sequence alignments (MSA) of a homologous protein family of interest;
    \item[$(ii)$] we train an energy-based generative model $P(S)\propto \exp\{-\mathcal{H}(S)\}$ recapitulating the statistical properties of the sequences in the MSA: here we use bmDCA \cite{Muntoni2021, Rosset2026adabmDCA2}, i.e.\ Direct Coupling Analysis based on Boltzmann machine learning, cf.~Materials and Methods:
    \begin{equation}
        \mathcal{H}(S) = -\sum_{i<j} J_{ij}(s_i,s_j) - \sum_{i} h_i(s_i)\ ;
    \end{equation}
    \item[$(iii)$] we use the inferred sequence landscape $\mathcal{H}(S)$ as a proxy for (negative) fitness to simulate protein evolution. To do so, we rely on a stochastic interplay between mutations at the nucleotide level, and amino-acid insertions and deletions, since this reflects better natural sequence evolution, and leads to a better quality of the simulated sequences in comparison with experimental results \cite{DiBari2024};
\end{enumerate}
Details of this simulation framework are given in Materials and Methods and the original publication \cite{DiBari2024}.
It allows us to generate evolutionary trajectories and collect data triples comprising a starting sequence $S_\mathrm{start}$, a temporal midpoint $S_\mathrm{mid}$ and a final sequence $S_\mathrm{end}$. More precisely, we construct $S_\mathrm{mid}$ as the exact intermediate in time, such that the evolutionary timescale $\tau$ elapsed between $S_\mathrm{start}$ and $S_\mathrm{mid}$ equals the time elapsed between $S_\mathrm{mid}$ and $S_\mathrm{end}$. We sample these trajectory segments across a logarithmic span of intervals, ranging from $\tau = 10^1$ to $\tau = 10^6$ Monte Carlo (MC) steps (which equal the attempted mutation, insertion, or deletion events). The upper bound of this range of MC steps corresponds approximately to the mixing time, representing a regime where all three sequences become effectively decorrelated, and any reconstruction of the intermediate based on the starting and end sequences becomes impossible (cf.~Materials and Methods). Within this ground-truth framework, we benchmark three distinct approaches for predicting $S_\mathrm{mid}$ given the boundary conditions $S_\mathrm{start}$ and $S_\mathrm{end}$, using the following partially complementary criteria:
\begin{itemize}
    \item the \textit{accuracy} of the predicted evolutionary intermediate is measured as the sequence identity to the simulated ground truth $S_\mathrm{mid}$; this is a standard measure for sequence reconstruction \cite{Williams2006, DeLeonardis2025};
    \item the Hamming \textit{distance of the reconstructed sequences from the starting sequence} $d(S_\mathrm{start},S_\mathrm{mid})$ measures the number of substituted sites between the initial and the reconstructed sequence, and can be compared to the same distance in the ground truth;
    \item the \textit{statistical energy} $\mathcal{H}(S_\mathrm{mid})$ assesses the quality of the reconstructed sequence within the original generative model, which can be seen as a ground truth quality score of sequences within our framework. This statistical energy can be compared to the simulated ground truth sequences. If too high, \textit{i.e.}\ reconstructed sequences have low generative probability, these sequences are expected to be of bad quality and do not correspond to functional sequences. If, on the contrary, they are lower than the ground truth sequence energies, reconstructed sequences are expected to be atypical in their statistical properties, and thereby influenced by artifacts of the inference procedure \cite{Russ2020, Levy2017}.
\end{itemize}
It should be noted that each of these quantities measures a different feature of the reconstructed sequences, therefore, concentrating on one quality measure alone might lead to a biased view on the reconstruction problem.

\subsubsection{Direct paths as the naive baseline.}
If we had knowledge only of the initial and final sequences $S_\mathrm{start}$ and $S_\mathrm{end}$, the most natural guess would be that the intermediate sequence $S_\mathrm{mid}$ would be approximately located in the middle of any of the direct paths between the endpoints \cite{Poelwijk2019Epistasis, Weinreich2006}, \textit{i.e.}\ in each polymorphic position, any of the two variant amino acids would be present with equal probability $p=0.5$, while monomorphic positions would also be assumed unmutated in $S_\mathrm{mid}$. For very short time scales, \textit{i.e.}\ only few mutations between the start and end sequences, we would expect this to be the best one can do: most observable mutations are expected to be close to neutral, and double mutations in the same position are extremely unlikely such that polymorphisms between start and end sequences correspond to single mutation events.

\begin{figure*}[!t]
    \centering
    \includegraphics[width=\textwidth]{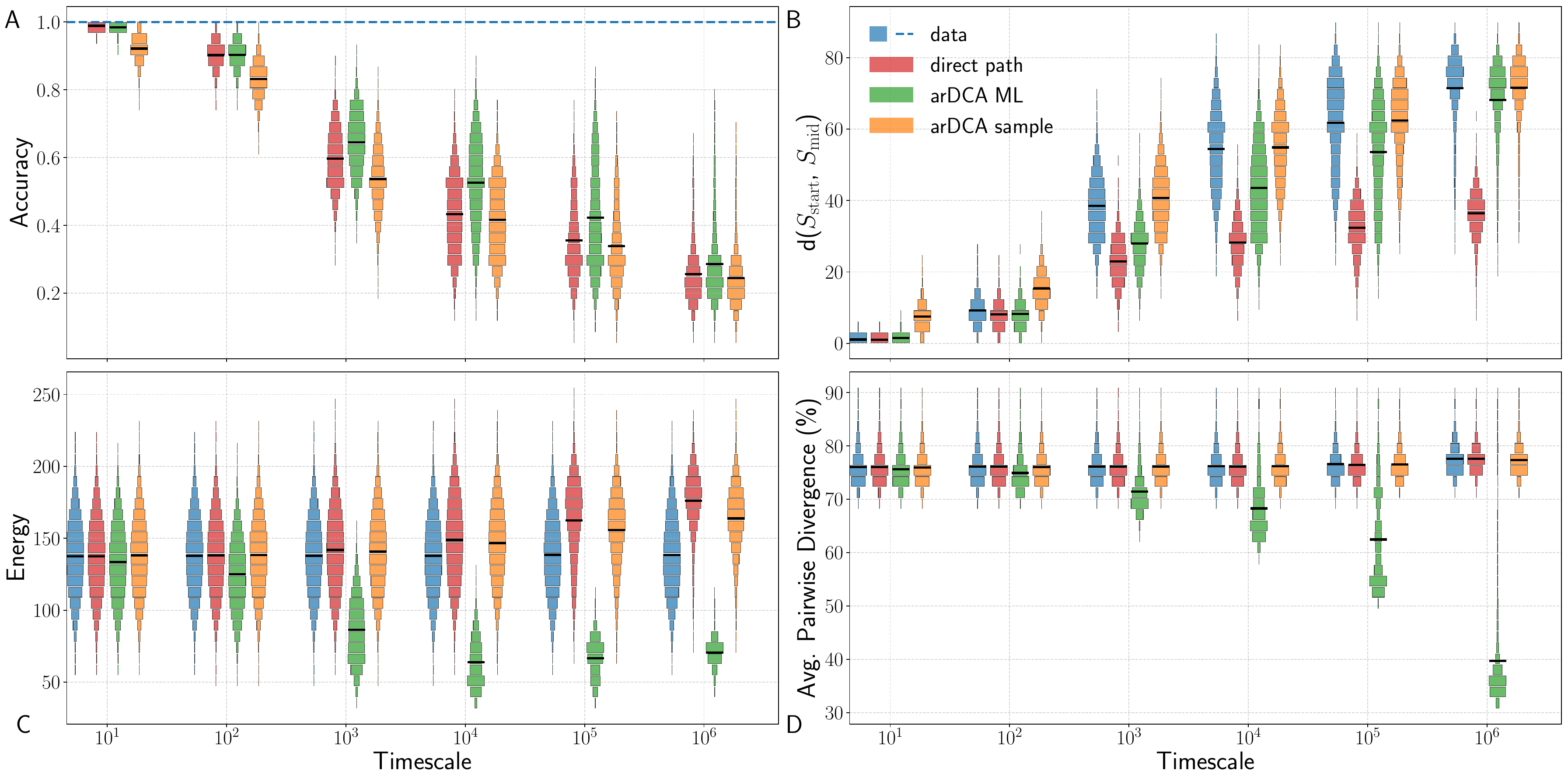}
    \caption{\textbf{Statistical comparison of prediction methods.} \textbf{(A)} Reconstruction accuracy of the three methods, with the blue line indicating perfect reconstruction (accuracy = 1). \textbf{(B)} Distribution of sequence divergence (Hamming distance) $d(S_\mathrm{start},S_\mathrm{mid})$ for the simulated ground truth (blue) compared to the three predictive approaches: arDCA sampling, direct path, and greedy arDCA (ML). \textbf{(C)} Distribution of sequence energies, as evaluated by the ground-truth Potts model. \textbf{(D)} Average pairwise sequence divergence (\%). Notably, at long evolutionary timescales, the greedy ML method tends to generate highly similar sequences therefore concentrated in a narrow region of the sequence space, independently of the specific endpoints used.}
    \label{fig:methods_comparison}
\end{figure*}

Fig.~\ref{fig:methods_comparison} confirms this expectation. On short time scales reconstruction accuracies are high, and distances to the start sequence and statistical energies are comparable to the values observed in the ground truth sequences. This changes for larger timescales from about $\tau\simeq 10^3$ MC steps. Although the reconstruction accuracy remains quite similar to the other methods discussed below, the sequences become substantially closer to the starting sequence than the ground truth, and the statistical energies $\mathcal{H}$ higher. The first observation reflects that, on these time scales, multiple mutations per site are no longer rare, and the intermediate sequence has amino acids in some positions which differ from both the starting and ending sequences. The second observation results from neglecting epistasis between mutations: on long time scales, the order of the mutations is no longer arbitrary, and the epistatic couplings between mutations rule out certain combinations, therefore giving higher energies in the generative DCA model. The random direct paths force the sequences to transition through high-energy barriers that a realistic evolutionary trajectory would circumvent \cite{Poelwijk2019Epistasis, Weinreich2006, Mauri2023}; thus the reconstructed sequences become biologically implausible compared to the ground truth.

\subsubsection{Maximum likelihood reconstruction via greedy arDCA.}
To improve beyond direct paths, we need to capture the missing effects, \textit{i.e.}\ multiple mutations and epistatic interactions between mutations. We can do so in a data-driven way by taking benefit from a large amount of simulated trajectories with known intermediate sequences. To this end, we train an autoregressive DCA model (arDCA, cf.~Materials and Methods for details) \cite{Trinquier2021} on trajectory triplets ordered as $(S_\mathrm{start}, S_\mathrm{end}, S_\mathrm{mid})$. This specific architecture allows the model to explicitly calculate the conditional distribution $P(S_\mathrm{mid} \mid S_\mathrm{start}, S_\mathrm{end})$ of the intermediate sequence given the boundary constraints. For reconstruction, we employ a deterministic maximum-likelihood prediction, often termed \textit{greedy decoding}. Rather than a global maximization, this method constructs $S_\mathrm{mid}$ autoregressively: proceeding site by site from the first to the last position, we select the amino acid with the highest probability, conditioned on the two endpoints and the previously reconstructed residues (cf.~Materials and Methods for the construction of training/validation/test sets for the arDCA-based reconstruction protocols).

As can be seen in Fig.~\ref{fig:methods_comparison}, this method maximizes the reconstruction accuracy, even if only by a relatively small margin when compared to the naive direct-path reconstruction. This maximization comes at a cost: it generates \textit{atypical} sequences that are statistically distinguishable from the simulated ground-truth ones.

The greedy reconstructions exhibit energy scores that are systematically lower compared to ground-truth trajectories. Moreover, they fail to reproduce the correct distribution of Hamming distances from the boundary sequences, placing the reconstructed intermediates in a geometric regime distinct from that of the simulated data. Furthermore, as the evolutionary timescales approach the mixing time and correlations between endpoints vanish, the greedy strategy decouples from the specific boundary conditions. In this regime, rather than recovering a diverse ensemble, the method tends to generate sequences that cluster densely in a specific region of the sequence space, exhibiting high similarity to each other independently of the specific endpoints used, thus limiting its utility for sampling realistic trajectories (Fig.~\ref{fig:methods_comparison}D).

\subsubsection{Generative sampling from arDCA.}
The third approach involves sampling directly from the conditional probability distribution $P(S_\mathrm{mid} \mid S_\mathrm{start}, S_\mathrm{end})$ learned by the arDCA model; sampling is easily achievable position by position thanks to the autoregressive architecture (cf.~Materials and Methods).

In contrast to the greedy approach, sampling reintroduces the stochasticity needed to generate \textit{typical} intermediates, rather than selecting an atypical sequence that is merely geometrically close to the true one. As shown again in Fig.~\ref{fig:methods_comparison}, the sampled intermediates reproduce the simulated ground-truth distributions of sequence divergence remarkably well across timescales. Their energy distribution overlaps closely with the simulated data at short timescales; at longer timescales, the sampled energies increase moderately, but they remain substantially closer to the ground-truth distribution than those obtained with either direct paths or maximum-likelihood decoding. Although sampling sacrifices a marginal amount of residue-level accuracy compared to the maximum-likelihood prediction, it better reconstructs the global statistical features of the evolutionary path. The only minor limitation arises at extremely short timescales, where the intrinsic variance of the model may introduce slightly more diversity than the other methods, a discrepancy that can be effectively mitigated by reducing the sampling temperature. However, for intermediate and long timescales, where navigating the energy landscape becomes critical, sampling from the learned conditional distribution provides the most realistic strategy for reconstructing evolutionary paths (Fig.~\ref{fig:methods_comparison}).

Note that the loss of accuracy when sampling instead of using maximum-likelihood reconstruction is an almost trivial statistical effect. In a sufficiently smooth, high-dimensional probability distribution, a typical sampled data point (here representing the ground truth) is closer to the maximum-probability point (representing the maximum-likelihood reconstruction) than to another typical, but independently sampled data point (representing reconstruction by sampling). The likelihood maximum remains still an atypical point, and the lack in accuracy by sampling is not a defect of the ``reconstruction'' algorithm, but an expression of the intrinsic reconstruction uncertainty resulting from the underlying stochastic dynamics.

To summarize, we compared several algorithms for reconstructing evolutionary intermediates between given start and end sequences. A generative model trained on simulated trajectories outperforms random direct-path interpolation at intermediate and long timescales, where recurrent mutations and epistasis violate the assumptions of the naive baseline. Maximum-likelihood reconstruction gives the highest residue-level accuracy, but at the cost of producing atypical intermediates with unrealistically low statistical energies. By contrast, sampling from the learned conditional distribution yields more typical, ``natural-like'' intermediates and reflects the intrinsic stochasticity of evolution. Thus, the appropriate goal is not to infer a single sequence, but to reconstruct an ensemble of plausible intermediates and their statistical properties. The same comparison for $\beta$-lactamase and the RR domain is reported in the \textit{Supplementary Appendix}, Section S6 and Figs.~S3--S4.

\subsection{Limits of prediction: crossing barriers and information loss}

The simplified setting of our work, where we reconstruct the temporal midpoint between an ancestral and a descendant sequence, allows us to gain clear and fundamental insight into the problem of evolutionary reconstruction. In a rugged sequence landscape, evolutionary trajectories do not explore all regions in the same way \cite{DiBari2024, Rossi2025, deVisserKrug2014}. This heterogeneity directly affects the reconstructability of evolutionary intermediates. In particular, our framework makes it possible to identify regimes in which information from one or both endpoint sequences is progressively lost, depending on the part of the sequence landscape explored by the trajectory.

\subsubsection{Information asymmetry: basins vs.\ entropic plateaus.}

The sequence landscape explored by the evolutionary dynamics is topologically heterogeneous \cite{DiBari2024, Rossi2025, deVisserKrug2014}.
In the terminology used here, we distinguish between constrained basins and permissive plateaus within the accessible part of the landscape.
Constrained basins correspond to low-energy regions, which in the landscape analogy used here play the role of local fitness peaks.
In these regions, sequences are mutationally entrenched: only a small subset of mutations is compatible with the surrounding sequence context, while many possible changes are strongly deleterious \cite{Vigue2022}.
As a consequence, evolution is more constrained and divergence from the current sequence is slower.
Permissive plateaus correspond to regions that typically have higher statistical energy than the basins, but are still accessible to the evolutionary dynamics.
In these regions, many more amino-acid states remain compatible with the sequence context, so sequences have a larger pool of close-to-neutral mutations and can explore sequence space more rapidly over comparable timescales \cite{DiBari2024, Rossi2025}.
These permissive plateaus should not be confused with strongly deleterious high-energy regions outside the evolutionarily accessible sequence space, which correspond to unfit or non-functional proteins.
Such regions are not explored by the evolutionary dynamics and are not the focus of this work (\textit{Supplementary Appendix}, Section S5.2 and Fig.~S1).

To quantify such topological differences, we use the average Context-Dependent Entropy ($\overline{\mathrm{CDE}}$) as a metric of sequence mutability \cite{Vigue2022, DiBari2024}, defined as:
\begin{eqnarray}
        \overline{\mathrm{CDE}} & = & \frac 1L \sum_{i=1}^L \mathrm{CDE}_i \nonumber \\
        \mathrm{CDE}_i & = &  - \sum_{s} P(s_i=s\, |\, S_{\setminus i} ) \log P(s_i=s\, |\, S_{\setminus i} )
        \label{eqn:CDE}
\end{eqnarray}
where $S_{\setminus i} = (s_1,\ldots,s_{i-1},s_{i+1},\ldots,s_L)$ denotes the sequence context of amino acid $s_i$, i.e.\ the entire amino-acid sequence but with position $i$ removed. The quantity $\mathrm{CDE}_i$ is the entropy of the conditional probability distribution of one site given the rest of the sequence. It therefore measures how many amino-acid states remain compatible with the current sequence context at that position, providing a local estimate of the pool of close-to-neutral mutations available to the sequence. When averaged over all sites, $\overline{\mathrm{CDE}}$ gives a global measure of sequence mutability in the landscape. Low $\overline{\mathrm{CDE}}$ corresponds to a constrained, low-energy state (basin), where only few mutations are tolerated, while high $\overline{\mathrm{CDE}}$ corresponds to a more permissive, high-entropy and higher-energy state (plateau). More details are given in \textit{Materials and Methods}.

We investigated the reconstructability of the intermediate sequence $S_\mathrm{mid}$ during evolutionary transitions between these topological regimes. To this end, we divided trajectory triplets from the test set into two classes: transitions from a basin to a plateau (low $\overline{\mathrm{CDE}} \to$ high $\overline{\mathrm{CDE}}$) and transitions in the opposite direction (high $\overline{\mathrm{CDE}} \to$ low $\overline{\mathrm{CDE}}$). Low- and high-$\overline{\mathrm{CDE}}$ sequences were defined using the 40th and 60th percentiles of the $\overline{\mathrm{CDE}}$ distribution computed from an equilibrium sample of the original generative model (the corresponding distributions and thresholds are shown in the \textit{Supplementary Appendix}, Section S5.3 and Fig.~S2). These thresholds provide a simple separation between constrained and permissive regions without making the classification too stringent. At the same time, the gap between the two thresholds avoids classifying as a transition a trajectory whose $\overline{\mathrm{CDE}}$ only fluctuates slightly around the median of the distribution. Depending on the protein family, the shortest timescales ($\tau = 10^1$, $\tau = 10^2$, and also $\tau = 10^3$ in the case of beta-lactamase) are excluded from this analysis, since they are generally insufficient for escaping from a basin or relaxing into one (\textit{Supplementary Appendix} Section S7, Tab.~S2, S4 and S7).

We then performed a data ablation study comparing the accuracy of the autoregressive prediction conditioned on both endpoints, $S_\mathrm{start}$ and $S_\mathrm{end}$, with predictions conditioned on only one endpoint. In this analysis, reconstruction accuracy is used as a proxy for how much information a given boundary provides about the intermediate sequence. By comparing the prediction obtained with both endpoints to the predictions obtained after removing one endpoint at a time, we can identify which boundary is most important for reconstructing $S_\mathrm{mid}$.

The results in Table~\ref{tab:low-highCDE} reveal a clear asymmetry. While evaluating the model on the complete, unfiltered test sets confirms that conditioning on both endpoints generally maximizes reconstruction accuracy (see the \textit{Supplementary Appendix}, Section S7 and Tables S1, S3, and S6), isolating trajectories that move between different topological regimes shows that information about the evolutionary intermediate sequence is not distributed equally between the two boundaries. In both cases, the boundary sequence of low $\overline{\mathrm{CDE}}$ carries more information about the intermediate sequence, and no or small information is added when adding the high-$\overline{\mathrm{CDE}}$ boundary in the conditioning. This suggests that the anchor for prediction is the sequence located in the constrained basin, the intuitive reason being that the evolutionary path spends more time close to the more constrained sequence. These findings were confirmed for $\beta$-lactamase and the RR domain; the corresponding results are reported in the \textit{Supplementary Appendix}, Section S7.2--S7.3 and Tables S5 and S8.

\begin{table}[!t]
\caption{\textbf{Endpoint ablation in basin--plateau transitions.} Reconstruction accuracy of autoregressive models conditioned on both endpoints, on $S_\mathrm{start}$ only, or on $S_\mathrm{end}$ only, for transitions between low- and high-$\mathrm{\overline{CDE}}$ regions of the landscape. In the current case (Chorismate Mutase) timescales $10^1$ and $10^2$ are omitted because they are insufficient for escaping from or relaxing into a basin. Values shown in bold correspond to the conditioning regimes with the highest reconstruction accuracy in each transition class, namely \emph{both endpoints} and \emph{start only} for Basin $\to$ Plateau transitions, and \emph{both endpoints} and \emph{end only} for Plateau $\to$ Basin transitions. The emerging pattern is that, when the trajectory connects a low- and a high-CDE region, the informative boundary is consistently the low-CDE endpoint. In some cases, incorporating the plateau endpoint in the joint prediction is not beneficial and can even slightly degrade reconstruction performance relative to conditioning on the basin endpoint alone. At the longest timescales, where triplets approach a near-decorrelated regime, these differences become negligible, as expected.
\label{tab:low-highCDE}}

\begin{tabular*}{\columnwidth}{@{\extracolsep\fill}lcccc@{}}
\toprule
\textbf{Conditioning} & \multicolumn{4}{c}{\textbf{Timescale ($\tau$)}} \\
\midrule
& $10^3$ & $10^4$ & $10^5$ & $10^6$ \\
\midrule
\multicolumn{5}{@{}l}{\textbf{Basin $\to$ Plateau} (Low $\overline{\mathrm{CDE}} \to$ High $\overline{\mathrm{CDE}}$)} \\
Both endpoints & \textbf{0.6403} & \textbf{0.5375} & \textbf{0.4047} & \textbf{0.2787} \\
Start only     & \textbf{0.6132} & \textbf{0.5361} & \textbf{0.4066} & \textbf{0.2806} \\
End only       & 0.6060 & 0.4845 & 0.3290 & 0.2754 \\
\midrule
\multicolumn{5}{@{}l}{\textbf{Plateau $\to$ Basin} (High $\overline{\mathrm{CDE}} \to$ Low $\overline{\mathrm{CDE}}$)} \\
Both endpoints & \textbf{0.6375} & \textbf{0.5390} & \textbf{0.4008} & \textbf{0.2753} \\
Start only     & 0.6020 & 0.4819 & 0.3236 & 0.2745 \\
End only       & \textbf{0.6133} & \textbf{0.5403} & \textbf{0.4089} & \textbf{0.2774} \\
\bottomrule
\end{tabular*}

\end{table}

\subsubsection{Information loss through high-CDE excursions.}

The previous analysis shows that low- and high-CDE regions do not contribute equally to the reconstruction of evolutionary intermediates. We now ask a related question: when a trajectory starts and ends in constrained low-CDE regions, can the intermediate still be reconstructed if the path crosses a high-CDE region in between? In the landscape picture, this corresponds to a ``barrier-crossing'' scenario: the two endpoints lie in basins, while the temporal midpoint is located on a mutationally permissive ``barrier'' separating constrained regions.

To test this, we classified the test-set triplets using the same low- and high-CDE thresholds defined above. We then compared two classes of triplets. In the first class, all three observed sequences have low $\overline{\mathrm{CDE}}$: $S_\mathrm{start}$, $S_\mathrm{mid}$, and $S_\mathrm{end}$ all belong to the constrained regime. These triplets correspond to trajectories that remain in a low-mutability region at the three sampled time points. In the second class, $S_\mathrm{start}$ and $S_\mathrm{end}$ have low $\overline{\mathrm{CDE}}$, whereas $S_\mathrm{mid}$ shows high $\overline{\mathrm{CDE}}$. These triplets therefore describe trajectories that start and end in constrained basins, but pass through a high-CDE barrier at the midpoint; the corresponding classification and sample sizes are reported in the \textit{Supplementary Appendix}, Section S8 and Tables S9--S13.
The results of this comparison show a clear difference between the two classes. When the three observed sequences remain in the constrained regime, maximum-likelihood reconstruction is consistently more accurate, cf.~Table~\ref{tab:cde_barrier_accuracy}. When the trajectory crosses a high mutability region, the accuracy decreases substantially. In the specific case of Chorismate Mutase, such barrier-crossing events are very rare at the shortest timescales, because the dynamics has not had enough time to complete such an excursion. For this reason, the comparison becomes meaningful only from the timescales where these events are sufficiently represented in the data. In the present dataset, the ML accuracy decreases from $0.664$ to $0.491$ at $\tau=10^4$, from $0.610$ to $0.362$ at $\tau=10^5$, and from $0.347$ to $0.284$ at $\tau=10^6$ (Table~\ref{tab:cde_barrier_accuracy}). The same comparison was performed for $\beta$-lactamase and the RR domain, and the corresponding results are reported in the \textit{Supplementary Appendix}, Sections S8.2--S8.3 and Tables S10--S13.

This makes evident that the loss of predictability is not only due to the endpoints being poorly constrained, as in the previous analysis. Here, the endpoints are both low-CDE sequences. The difficulty instead comes from the high-CDE region crossed by the trajectory: because this region is highly mutable and has large entropy, the number of possible routes connecting the same endpoints becomes very large. As a consequence, the specific midpoint followed by one trajectory becomes difficult to infer from the endpoints alone.

This interpretation is also robust to two possible concerns. The classification is defined only from the three sampled points along each trajectory. At short timescales, this is likely to correspond to trajectories that remain inside a constrained basin. At longer timescales, however, unobserved excursions may occur between the sampled points. This means that some trajectories assigned to the constrained class may in fact have crossed a high-CDE region between two observations. Such hidden excursions would make the constrained class more heterogeneous and reduce its reconstruction accuracy. Therefore, the observed gap between the two classes is likely to underestimate, rather than overestimate, the effect of barrier crossing.

A second point to consider is the possible role of endpoint divergence. In general, intermediates between closely related endpoints are easier to reconstruct than intermediates between more distant endpoints. Moreover, trajectories that cross a high-CDE barrier also tend to connect endpoints that are farther apart. Therefore, the lower accuracy observed for barrier-crossing trajectories could in principle be explained simply by their larger endpoint divergence. However, this is not sufficient to account for the effect observed here. As shown in the \textit{Supplementary Appendix}, Section S8 and Figs.~S5--S7, the reduction in accuracy persists when the two trajectory classes are compared at fixed endpoint divergence. Thus, even among trajectories whose endpoints are separated by the same Hamming distance, barrier-crossing paths remain harder to reconstruct than constrained paths. The drop in reconstruction accuracy is therefore not merely a consequence of comparing closer and more distant endpoints, but specifically reflects the loss of path information associated with crossing a high-CDE barrier.

This result complements the endpoint-ablation analysis. We showed that when one boundary sequence lies in a high-CDE plateau, it carries little or no information about the intermediate compared with a low-CDE boundary. We see a corresponding effect along the trajectory itself: crossing a high-CDE region between two constrained endpoints reduces the reconstructability of the midpoint. High-CDE regions of the sequence landscape cause information loss. They make evolution faster and more permissive, but at the same time they weaken the memory of the particular mutational path followed by the trajectory.

\begin{table}[!t]
\caption{\textbf{Effect of barrier crossing on ML reconstruction accuracy.}
Triplets are split into two classes using the same low- and high-CDE thresholds defined above. In the first class, the start, midpoint, and endpoint all belong to the low-CDE regime, corresponding to constrained regions of the landscape. In the second class, the two endpoints belong to the low-CDE regime, whereas the midpoint belongs to the high-CDE regime. These triplets therefore correspond to trajectories that start and end in constrained regions, but cross a permissive high-CDE barrier at the midpoint. In the specific case of Chorismate Mutase, barrier-crossing triplets are essentially absent at short timescales, because the trajectory has not had enough time to leave a basin and return to a low-CDE state through a high-CDE region; this is why no barrier-crossing accuracy is reported before $\tau=10^4$.
}
\label{tab:cde_barrier_accuracy}

\begin{tabular*}{\columnwidth}{@{\extracolsep\fill}lcccccc@{\extracolsep\fill}}
\toprule
\textbf{Trajectory class} &
\multicolumn{6}{c}{\textbf{Evolutionary timescale ($\tau$)}} \\
& $10^1$ & $10^2$ & $10^3$ & $10^4$ & $10^5$ & $10^6$ \\
\midrule
\textbf{Constrained path} &
0.987 & 0.923 & 0.728 & 0.664 & 0.610 & 0.347 \\
\textbf{Barrier crossing} &
-- & -- & -- & 0.491 & 0.362 & 0.284 \\
\bottomrule
\end{tabular*}
\end{table}

\subsection{Towards real-case applications: inferring evolutionary time from sequence data}

\subsubsection{Disentangling distance and time via landscape topology.}

In biological applications, we do not know the exact evolutionary time $\tau$ elapsed between two sequences. In addition, it is generally difficult to know how the timescale of real evolution should be matched to the Monte Carlo time used in simulated evolution. By contrast, sequence divergence is directly measurable, here as the Hamming distance between aligned sequences. This creates a practical difficulty: the time-parameterized model requires a timescale as input, but this quantity is not available in real applications.

\begin{figure}[!t]
    \centering
    \includegraphics[width=\columnwidth]{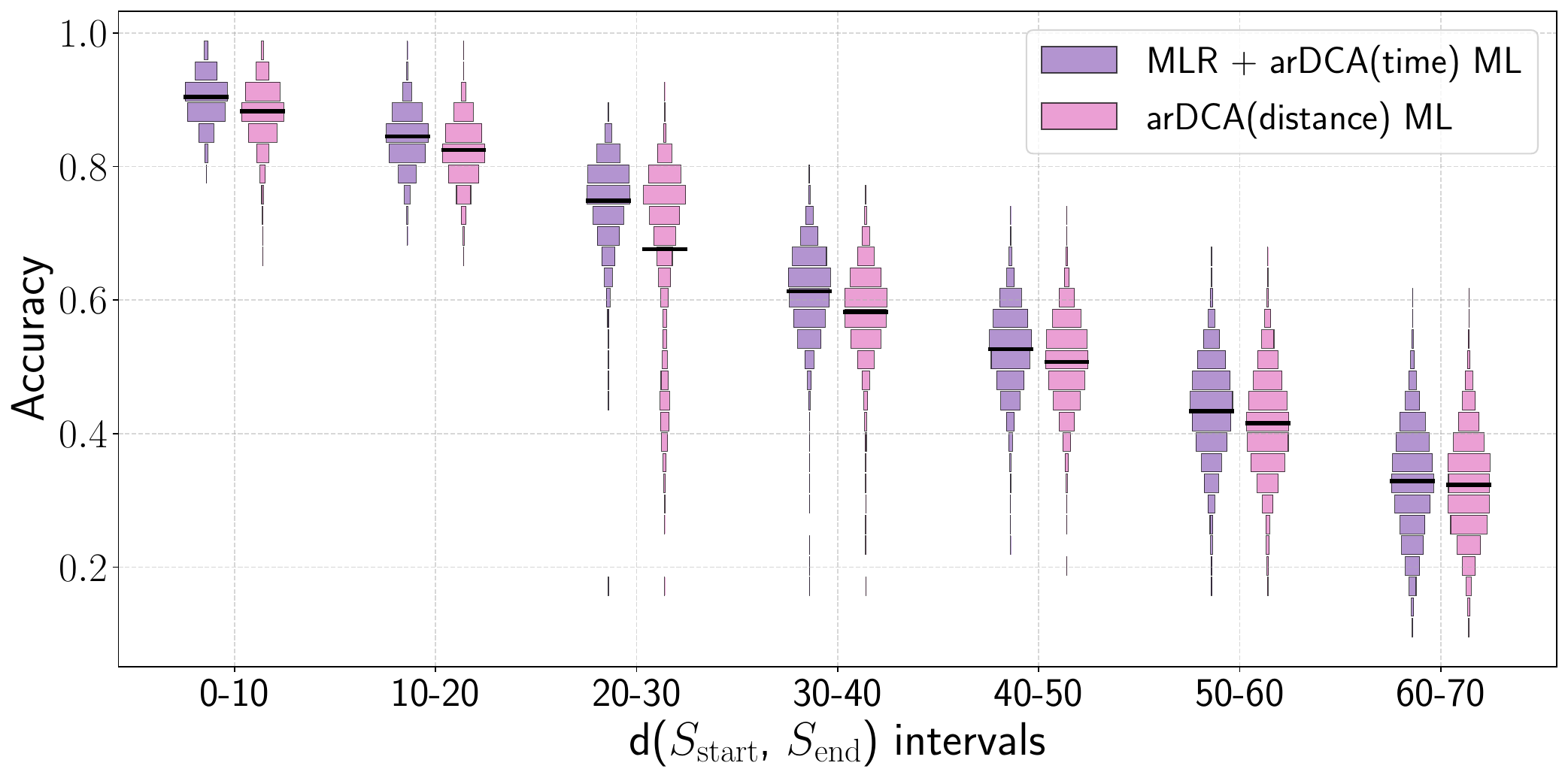}
    \caption{\textbf{Distance-based conditioning fails to capture the correct intermediate position in sequence space.} Reconstruction accuracy with ML greedy method for different intervals of endpoint divergence $d(S_\mathrm{start},S_\mathrm{end})$, using the same divergence bins as defined in the main text. For each bin, we compare a time-parameterized autoregressive model, $\mathrm{arDCA}(\mathrm{time})$, in which the evolutionary timescale is first inferred by multinomial logistic regression (MLR) and the corresponding time-trained model is then used for prediction; second, a distance-conditioned autoregressive model, $\mathrm{arDCA}(\mathrm{distance})$, trained directly on triplets grouped by endpoint divergence intervals. The time-parameterized approach based on MLR-inferred timescales yields systematically higher accuracy. For readability, adjacent endpoint-divergence bins are grouped in pairs, reducing the number of labels on the $x$-axis.}
    \label{fig:real_cases_hist}
\end{figure}

The simplest choice would be to train the model using sequence divergence rather than evolutionary time. In practice, for Chorismate Mutase, we implemented this by grouping trajectory triplets into divergence intervals between $S_\mathrm{start}$ and $S_\mathrm{end}$, for example $[0,5], [5,10], \dots, [65,70]$ mutations (see Materials and Methods and \textit{Supplementary Appendix}, Section S9.1 for further details). However, this distance-based modeling does not provide the best reconstruction accuracy (Fig.~\ref{fig:real_cases_hist}) or the best-calibrated statistical properties of intermediate sequences (see the \textit{Supplementary Appendix}, Section S9.3 and Fig.~S9). A likely reason is that the same endpoint divergence can arise from different types of trajectories, depending on the position of the starting sequence in the landscape. Trajectories starting from low-$\overline{\mathrm{CDE}}$ sequences tend to remain constrained near the starting point and therefore require longer times to reach a given divergence. By contrast, trajectories starting from high-$\overline{\mathrm{CDE}}$ sequences diverge more rapidly, reaching the same sequence distance in shorter time.

This asymmetry makes the distance-conditioned problem intrinsically harder. In the time-parameterized model, the evolutionary timescale is given explicitly, while the endpoint divergence can be measured directly from the two conditioning sequences. In the distance-parameterized model, the endpoint divergence is given by construction, but the elapsed time remains hidden. Pooling trajectories only by endpoint divergence can therefore mix paths with different timescales and different dynamical regimes, making the prediction less well calibrated.

This shows that \textit{sequence divergence} is not an accurate proxy for simulated evolutionary time: the same divergence can correspond to different elapsed times depending on the local topology of the landscape. The evolutionary time between the boundary sequences is an important input for accurate reconstruction of the intermediate sequence.

\subsubsection{Recovering evolutionary time via topological features.}
\begin{figure}
    \centering
    \includegraphics[width=1\linewidth]{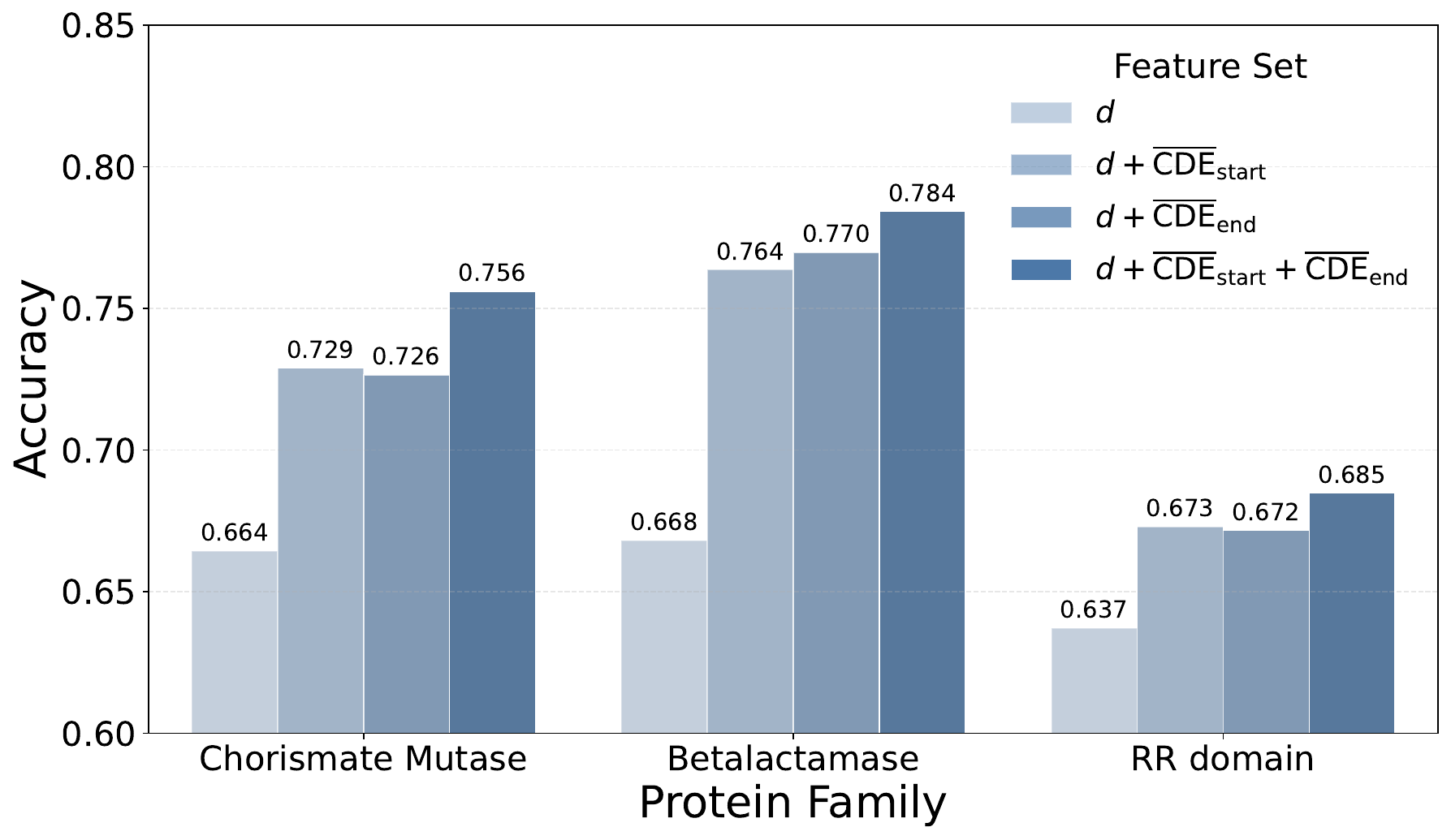}
    \caption{Accuracy of multinomial logistic regression models trained to predict the evolutionary timescale from different combinations of observables. Here, $d$ denotes the Hamming distance between the two endpoint sequences. Using both $\overline{\mathrm{CDE}}_\mathrm{start}$ and $\overline{\mathrm{CDE}}_\mathrm{end}$ together with $d$ improves performance over divergence information alone.}
    \label{fig:MLR}
\end{figure}

To overcome this problem, we propose to invert the observation that the sequence divergence in a given evolutionary time depends on the sequence mutability, as measured by the average CDE. We propose inferring the \textit{a priori} unknown evolutionary time $\tau$ between two sequences $S_\mathrm{start}$ and $S_\mathrm{end}$ from the directly measurable sequence divergence (here represented via the Hamming distance between sequences) and the mutabilities of the two extremal sequences (represented by the two CDEs). We have observed that this can be achieved via a simple multinomial logistic regression. We also tested more detailed feature sets, including the binary mismatch vector between endpoints and residue-level CDE profiles, and observed only a modest gain over simpler CDE-based features (see the \textit{Supplementary Appendix}, Section S10.1 and Fig.~S10). In contrast, the inference becomes substantially less precise when only the Hamming distance alone or with a single CDE are used (Fig.~\ref{fig:MLR}).

We may now directly use the inferred time $\hat{\tau}$ in the time-parameterized generative models described above. As shown in Fig.~\ref{fig:real_cases_hist}, the statistical properties of the reconstructed sequences now match much more closely those of the simulated ground-truth data, even if the true time was unknown. This yields a fundamental implication for the application of data-driven models to real scenarios of path prediction and sampling: because sequence divergence and raw initial and final sequences alone are insufficient to infer evolutionary time, explicitly integrating topological landscape information (via CDE) is essential to estimate the evolutionary rate, identify the correct timescale, and thus accurately constrain the realistic position of the intermediate state within the sequence landscape.

\section{Discussion}

In this work, we used a controlled \textit{in silico} framework to study a difficult question that is usually inaccessible in real data: how much of an evolutionary intermediate sequence can be reconstructed from its two endpoints. Combining family-specific generative landscapes, nucleotide-level evolutionary simulations, and autoregressive conditional modeling \cite{DiBari2024, Cocco2018, Trinquier2021}, we were able to separate three issues that are often conflated: the statistical quality of a reconstructed sequence, its similarity to the true intermediate, and the intrinsic predictability of the evolutionary path itself.

\subsubsection{Beyond residue-level accuracy.}

Our first result highlights an important point for intermediate reconstruction: residue-level accuracy is only one criterion for evaluating a predicted sequence. A reconstruction may be close to the true midpoint in terms of sequence identity, while still having unrealistic properties in the landscape. For this reason, we propose to evaluate reconstructed intermediates also through complementary quantities, such as their distance from the endpoints and their statistical energy.

This distinction becomes crucial for both the direct-path baseline and greedy maximum-likelihood decoding. Direct paths can be reasonable at very short timescales, but fail at longer timescales because they ignore recurrent mutations and epistatic constraints. Greedy decoding improves pointwise accuracy, but produces atypical intermediates with distorted distances and unusually low energies.

These results reflect the stochastic nature of evolutionary trajectories. The true intermediate is only one realization among many possible paths connecting the same endpoints. The relevant object is therefore not necessarily a single optimal sequence, but an ensemble of plausible intermediates. Sampling from $P(S_\mathrm{mid}\mid S_\mathrm{start},S_\mathrm{end})$ better preserves these ensemble properties, even if it slightly reduces pointwise accuracy. Thus, assessing reconstructed intermediates requires asking not only how many residues are correctly recovered, but also whether the predictions have the statistical properties expected from realistic trajectories, and whether the reconstructed mutations are compatible with each other and with the rest of the sequence.

\subsubsection{Landscape topology limits prediction.}

Predictability is strongly influenced by the topology of the sequence landscape \cite{Rossi2025, deVisserKrug2014}. When the trajectory remains within constrained low-CDE regions, the endpoints retain substantial information about the intermediate, and accurate reconstruction can remain possible even at relatively long times. In these basin-like regions, evolution is slower and more constrained: only a limited set of mutations is compatible with the sequence context, and the possible routes between two endpoints are therefore less numerous.

The situation changes when the trajectory reaches a high-CDE region. Our results show that these regions act as an \textit{information horizon}: the memory of the specific evolutionary route is progressively weakened. If one endpoint lies on a high-CDE plateau, that endpoint carries substantially less information about the path than the endpoint located in a constrained basin. If the path itself crosses a high-CDE barrier between two low-CDE endpoints, the specific midpoint also becomes harder to reconstruct. This is because high-CDE regions are more permissive: many different mutational routes are possible. In addition, the barrier-crossing event itself is highly stochastic, and this contributes to erasing the memory of the particular route followed.

In this regime, the limitation is not a modeling problem nor a lack of training data. It is a basic property of stochastic evolution on a rugged landscape. This distinction between low-CDE and high-CDE regions is therefore important for moving toward real-case applications: it helps identify when reconstructing an evolutionary intermediate is feasible, and when the endpoints no longer contain enough information to do so reliably.

\subsubsection{From sequence divergence to evolutionary time.}

Our third main result concerns the transition from simulation to real applications. In natural data, the elapsed evolutionary time between two sequences is unknown, whereas sequence divergence is directly observable. However, our results show that training reconstruction models based on endpoint divergence is not the right solution. When the model is trained using evolutionary time, it receives the elapsed time explicitly, while the divergence between the endpoints is already contained in the two input sequences. When the model is trained using only endpoint divergence, instead, the elapsed time remains hidden. It therefore mixes trajectories that have the same distance but belong to different dynamical regimes. As a result, it cannot resolve these different regimes.

We show that this problem can be addressed by adding explicit topological information. In practice, combining endpoint divergence with mutability estimates of the two endpoint sequences allows us to infer the latent evolutionary timescale with good accuracy. Once this timescale is estimated, the corresponding generative model trained at that timescale produces more realistic intermediates. Realistic path reconstruction is therefore not impossible in the absence of explicit temporal information, but it cannot be achieved from endpoint divergence alone: the latent evolutionary time must first be inferred from the topology of the sequence space around the endpoint sequences.

\subsubsection{Conclusion and Future Perspectives.}

Taken together, our results provide some interesting answers to the question posed in this work. Intermediate evolutionary sequences can be reconstructed in a statistically meaningful way, but only within limited regimes. It is therefore important to ask not only which intermediates are plausible, but also whether the endpoints contain enough information to reconstruct them reliably.

More broadly, this work provides a framework for learning evolutionary paths from simulated trajectories in family-specific sequence landscapes. For more flexible and modular architectures, such as transformers, our results suggest that topological information associated with the endpoint sequences could provide an important conditioning signal, in line with recent efforts into deep models of protein evolution \cite{Koehl2026PEINT}. This information would help the model distinguish different dynamical regimes and could ultimately allow a single predictor to operate across multiple evolutionary timescales and divergence ranges.

\section{Materials and Methods}

\subsection{Sequence data and landscape inference}

The main dataset used in this study is the multiple sequence alignment (MSA) of the Chorismate Mutase protein family, previously investigated by \cite{Russ2020, Netti2026HighEntropy}. This family is used as the primary benchmark throughout the main text. To assess the generality of the proposed framework, we also applied the same analysis pipeline to two additional protein families, the response regulator domain (RR domain, PF00072) and $\beta$-lactamase (PF13354). The corresponding results, together with the details of MSA collection, preprocessing, and filtering for all protein families, are reported in the \textit{Supplementary Appendix}, Section S1, S6, S7 and S8.
To describe the epistatic sequence landscape of each protein family, we use Boltzmann Machine Direct Coupling Analysis (bmDCA) \cite{Cocco2018, Muntoni2021, Rosset2026adabmDCA2}, trained directly on the corresponding MSA. This generative statistical model assigns a statistical energy $\mathcal{H}(S)$ to each amino-acid sequence $S = (s_1,\dots,s_L)$, accounting both for the independent conservation of individual sites and for pairwise epistatic interactions between sites. The Hamiltonian is defined as
\begin{equation}
\mathcal{H}(S) = -\sum_{i<j}J_{ij}(s_{i},s_{j}) - \sum_{i}h_{i}(s_{i}) \, .
\end{equation}

The inferred bmDCA model has two roles in this work. First, it defines the artificial sequence landscape in which we simulate evolution and generate trajectory data. This landscape is inferred from the natural MSA and is used as a proxy for the unknown true fitness landscape. While artificial, it preserves key statistical features of the corresponding protein family and provides a controllable landscape that can be explicitly sampled.

Second, because the simulated trajectories are generated within this same landscape, the bmDCA model also provides the reference against which predictions are evaluated. In this controlled setting, a predicted intermediate can therefore be assessed by scoring it with the same model that generated the evolutionary dynamics (\textit{Supplementary Appendix} S2).

\subsection{Predicted mutational robustness and context-dependent entropy}

To quantify the mutability of a site within a given sequence background, we evaluate its Context-Dependent Entropy (CDE) \cite{Figliuzzi2016, Vigue2022, Poelwijk2016, RodriguezRivas2022}. Using the conditional probabilities assigned by the bmDCA model for all possible amino-acid states at site $i$, given a fixed sequence context $S_{\backslash i}$ (the full sequence excluding site $i$), the local CDE is defined as in Eqn.~\ref{eqn:CDE}.
The CDE provides an interpretable information-theoretic measure of local mutability. A high CDE at site $i$ indicates that the surrounding sequence context imposes weak constraints, so that several amino-acid states remain compatible with the local background. Conversely, a low CDE indicates that the site is strongly constrained by the current sequence context, reflecting epistatic entrenchment.

By averaging the CDE over all sites of a sequence, we obtain a global estimate of its overall mutability (Eqn.~\ref{eqn:CDE}).
This average quantity characterizes how constrained or permissive the genetic background is for evolutionary exploration starting from that specific point in sequence space. As shown in the \textit{Supplementary Appendix}, Section S5.2 and Fig.~S1, $\overline{\mathrm{CDE}}$ is strongly correlated with the statistical energy of the sequence: low-energy sequences, which lie in deep basins of the landscape, typically have low $\overline{\mathrm{CDE}}$, whereas higher-energy sequences in flatter, plateau-like regions tend to have higher $\overline{\mathrm{CDE}}$. Since $\overline{\mathrm{CDE}}$ measures mutability, this means that sequences in low-energy basins are generally more constrained and less mutable, while sequences in high-$\overline{\mathrm{CDE}}$ regions are more permissive to mutations. This directly affects the pace of evolutionary dynamics: low-$\overline{\mathrm{CDE}}$ backgrounds evolve more slowly, whereas high-$\overline{\mathrm{CDE}}$ backgrounds allow faster exploration of sequence space.

\subsection{Simulating evolution}

Evolutionary dynamics are then simulated in the inferred landscape as a stochastic Markov Chain Monte Carlo (MCMC) process, using the computational framework previously introduced by \cite{DiBari2024}. To make the simulations more biologically realistic, mutations are modeled at the nucleotide level rather than directly at the amino-acid level \cite{Goldman1994}. This allows the dynamics to account for both the constraints and the degeneracy of the genetic code.

The simulation relies on a mixed sampler combining Gibbs sampling for single-nucleotide substitutions and Metropolis sampling for codon-level insertions and deletions (indels). Indels are incorporated within the aligned framework by introducing gaps when a codon is deleted and by inserting the new nucleotide triplet into an existing gap during insertion events. This sampling procedure satisfies detailed balance and therefore ensures that the long-time dynamics correctly sample the equilibrium distribution corresponding to the inferred bmDCA model.

To generate the trajectory data, the simulation chains are initialized from an equilibrium sample of the ground-truth bmDCA model composed of 100,000 i.i.d.\ sequences. Starting from these sampled sequences, each chain is evolved for a large number ($10^7$) of steps, where each step corresponds to one attempted mutation or indel event. From these simulations, we extract ground-truth trajectory triplets $(S_\mathrm{start}, S_\mathrm{mid}, S_\mathrm{end})$ across a logarithmically spaced range of evolutionary timescales,
\[
\tau \in \{10^1, 10^2, 10^3, 10^4, 10^5, 10^6\} \, ,
\]
where $\tau$ denotes the evolutionary time between $S_\mathrm{start}$ and $S_\mathrm{mid}$, and equally between $S_\mathrm{mid}$ and $S_\mathrm{end}$. For each value of $\tau$, we constructed a timescale-specific dataset of trajectory triplets extracted from the simulations. Each timescale dataset contains 70,000 triplets for training, 10,000 for validation, and 20,000 for testing. All results presented in this work are evaluated on the corresponding test sets.

The upper bound of the timescale range was chosen to cover a broad dynamical regime, from short-time local evolution to long-time behavior in which sequences are expected to have substantially explored the landscape and become only weakly correlated with their initial states. In the limiting case of decorrelation, no mutual information remains between the endpoints, making reconstruction from boundary sequences alone intrinsically difficult.

Such long simulations also imply that some chains may occasionally return close to their original state, a phenomenon that can arise in long-run MCMC sampling but should not be interpreted as a realistic feature of natural evolution. This does not affect the purpose of our analysis, which is to study the predictability of intermediate sequences under controlled landscape dynamics, rather than to reproduce complete natural evolutionary histories. Nevertheless, this point should be kept in mind when interpreting trajectories at the longest timescales.

\subsection{Path inference and time intermediate reconstruction}

\subsubsection{Direct path inference}

As a naive baseline, we define a reconstruction procedure based on direct-path interpolation between the two boundary sequences. Intermediate sequences are sampled under an independent-site assumption: at each polymorphic site between $S_\mathrm{start}$ and $S_\mathrm{end}$, the two amino acids observed in the endpoints are assigned equal probability $p=0.5$, and the intermediate residue is sampled from this two-state distribution. At non-polymorphic sites, the same amino acid is kept in the intermediate sequence. This corresponds to sampling an intermediate sequence belonging to a random direct mutational path between the start and end sequences. As a consequence, this reconstruction depends only on the two endpoints and on the mutational distance separating them, without taking into account the topology of the sequence space that connects the two sequences.

\subsubsection{Autoregressive modeling}

As a data-driven reconstruction approach, we use arDCA to learn from the simulated trajectory triplets \cite{Trinquier2021}. The original data are triplets of the form
$(S_\mathrm{start}, S_\mathrm{mid}, S_\mathrm{end})$. Since our goal is to predict the intermediate sequence from the two endpoints, each triplet is reorganized in the order
$(S_\mathrm{start}, S_\mathrm{end}, S_\mathrm{mid})$.
With this ordering, the intermediate sequence is generated after the two endpoints and is therefore conditioned on both. For notational clarity, in the equations below we write the labels \textit{start}, \textit{end}, and \textit{mid} as superscripts, although they have the same meaning as in the text. The autoregressive factorization is then:

\begin{equation}
\begin{aligned}
&P(S^\mathrm{start}, S^\mathrm{end}, S^\mathrm{mid}) = \\
&\quad
\prod_{i=1}^{L}
P(s^\mathrm{start}_{i} \mid S^\mathrm{start}_{<i})
\prod_{i=1}^{L}
P(s^\mathrm{end}_{i} \mid S^\mathrm{start}, S^\mathrm{end}_{<i}) \\
&\quad \times
\prod_{i=1}^{L}
P(s^\mathrm{mid}_{i} \mid S^\mathrm{start}, S^\mathrm{end}, S^\mathrm{mid}_{<i}) \, .
\end{aligned}
\end{equation}

The last product is the quantity of interest, namely
$P(S^\mathrm{mid}\mid S^\mathrm{start},S^\mathrm{end})$. Thus, the intermediate sequence can be generated residue by residue, with each site depending on the previously generated residues of $S_\mathrm{mid}$ and on all sites of the two endpoint sequences. In the arDCA framework \cite{Trinquier2021}, the probability of each intermediate residue can be written as follows:

\begin{equation}
\begin{aligned}
& P(s^\mathrm{mid}_{i} \mid S^\mathrm{mid}_{< i}, S^\mathrm{start}, S^\mathrm{end})
= \frac{1}{z_i\!\left(S^\mathrm{mid}_{<i}, S^\mathrm{start}, S^\mathrm{end}\right)} \times \\
& \hspace{2.2cm} \times \exp\Biggl\{
h_i(s^\mathrm{mid}_{i}) + \sum_{j=1}^{i-1} J_{ij}(s^\mathrm{mid}_{i},s^\mathrm{mid}_{j}) +\\
& \hspace{2.2cm}
+ \sum_{j=1}^{L}\!\Bigl[
K_{ij}(s^\mathrm{mid}_{i},s^\mathrm{start}_{j})
+ M_{ij}(s^\mathrm{mid}_{i},s^\mathrm{end}_{j})
\Bigr]
\Biggr\}\, .
\end{aligned}
\end{equation}

with

\begin{equation}
\begin{aligned}
&z_i\!\left(S^\mathrm{mid}_{< i}, S^\mathrm{start}, S^\mathrm{end}\right)
= \\
&\hspace{2.1cm} \sum_{s^\mathrm{mid}_{i}} \exp\Biggl\{
h_i(s^\mathrm{mid}_{i}) + \sum_{j=1}^{i-1} J_{ij}(s^\mathrm{mid}_{i},s^\mathrm{mid}_{j}) + \\
&\hspace{2.1cm} + \sum_{j=1}^{L}\!\Bigl[
K_{ij}(s^\mathrm{mid}_{i},s^\mathrm{start}_{j})
+ M_{ij}(s^\mathrm{mid}_{i},s^\mathrm{end}_{j})
\Bigr]
\Biggr\}\, .
\end{aligned}
\end{equation}

Here $J$, $K$, $M$ represent the couplings between mid--mid, mid--start, and mid--end residues, respectively.
This expression highlights the two contributions entering the probability of each intermediate site: the dependence on the residues of $S_\mathrm{mid}$ already generated, and the dependence on all sites of the start and end sequences.

To accelerate training, we modify the standard arDCA optimization procedure by computing the likelihood only over the residues of the final block of the concatenated array, which here corresponds to $S_\mathrm{mid}$, instead of over all sites. This is equivalent to maximizing $P(S^\mathrm{mid}\mid S^\mathrm{start},S^\mathrm{end})$, which is the quantity needed for intermediate reconstruction.

A separate model is trained for each timescale using the corresponding training set. Each model is then used to predict the intermediate sequences of triplets belonging to the test set of the same timescale, that is, triplets that were never seen during training.

\subsubsection{Greedy maximum likelihood}

Once the arDCA model has been trained, different decoding strategies can be used to reconstruct the evolutionary intermediate. The most direct deterministic strategy is greedy maximum-likelihood decoding, in which the intermediate sequence is built site by site by selecting, at each step, the amino acid with maximum conditional probability given the boundary sequences and the residues of the intermediate already generated:
\begin{equation}
    s^\mathrm{mid,ML}_i = \arg\max_{s^\mathrm{mid}_i}
    P(s^\mathrm{mid}_i \mid S^\mathrm{mid}_{< i}, S^\mathrm{start}, S^\mathrm{end}) \, .
\end{equation}
This deterministic decoding procedure is applied sequentially from the first residue of the intermediate sequence to the last. It is important to note that this does not correspond to the absolute maximum of the full conditional distribution, but rather to the result of a greedy site-by-site maximization.

\subsubsection{Generative sampling}

Intermediate sequences can also be generated by direct sampling from the conditional distribution learned by the arDCA model,
\[
S^\mathrm{mid} \sim P(S^\mathrm{mid}|S^\mathrm{start},S^\mathrm{end}) \, .
\]
Sampling is performed autoregressively, residue by residue, using the same conditional probabilities defined above. In the present work, all samples drawn from the autoregressive models were generated at temperature $T=1$. Lowering the temperature from 1 toward 0 progressively makes the sampling procedure more deterministic, approaching greedy maximum-likelihood decoding in the zero-temperature limit. For each triplet in the test set, the generative prediction corresponds to one independent sample from the posterior distribution.

\subsubsection{Model conditioned to one endpoint}

For the ablation study in which the model is conditioned only on the starting point or only on the endpoint, the autoregressive framework is modified by removing one of the two conditioning sequences. The corresponding target distribution becomes $P(S^\mathrm{mid}|S^\mathrm{endpoint})$, written as a product of conditional probabilities of the form $P(s^\mathrm{mid}_{i} \mid S^\mathrm{mid}_{< i}, S^\mathrm{endpoint})$.

Training is then performed on pairs rather than triplets, namely $(S^\mathrm{start}, S^\mathrm{mid})$. Since the simulation framework is invariant under time reversal, training on pairs ordered forward or backward in time is equivalent. Apart from this modification, the training pipeline remains identical to the one described above.

\subsubsection{Distance-conditioned training}

For the distance-conditioned reconstruction analysis, we used the Chorismate Mutase trajectory data and grouped triplets according to the Hamming distance between the two endpoint sequences, $d(S_\mathrm{start},S_\mathrm{end})$, rather than according to the evolutionary timescale $\tau$. Training, validation, and test sets were handled separately: within each split, triplets from all timescales were merged and then assigned to endpoint-divergence intervals of width 5, namely $[0,5]$, $[5,10]$, $\dots$, $[65,70]$.

The upper bound of 70 was chosen to restrict the analysis to divergence intervals that remain informative about the progression of the dynamics. For Chorismate Mutase, the equilibrium pairwise Hamming distance between independently sampled sequences is slightly above this value. Up to this range, endpoint divergence still increases with evolutionary time, so different divergence intervals can be interpreted as distinct dynamical regimes. Beyond this point, the distance approaches saturation: trajectories generated at different timescales can fall into the same divergence range, making it difficult to separate evolutionary regimes (\textit{Supplementary Appendix}, Fig.~S8). Training distance-conditioned models beyond this range would therefore risk mixing trajectories with largely different evolutionary histories but similar endpoint distances, making the conditioning variable much less informative.

For each divergence interval, triplets were ordered as $(S_\mathrm{start},S_\mathrm{end},S_\mathrm{mid})$, as in the time-conditioned analysis, and a separate autoregressive model was trained to predict $S_\mathrm{mid}$ from the two endpoints. The only difference from the previous training procedure is therefore the grouping criterion: endpoint divergence rather than evolutionary timescale.

\subsection{Inference of evolutionary timescale}

The inference of the evolutionary timescale is formulated as a multiclass classification problem. Each triplet in the training set is assigned a label corresponding to its true timescale, and the task is then to predict this label from observable features derived from the two endpoint sequences.

Several sets of input features were considered. These include the Hamming distance between the endpoints, the full binary mismatch vector indicating for each site whether the two endpoint amino acids are identical or different, the average CDE values of the start and end sequences, and the full site-resolved CDE vectors of the two endpoints. The use of full vector features or of the corresponding sequence-averaged quantities led to comparable results.
All classifiers were trained on 30,000 of the training triplets, using an 80/20\% split for training and validation to obtain the benchmark results reported in Fig.~\ref{fig:MLR}, and were subsequently applied to the corresponding test sets.

\section{Data availability}
All code and data to reproduce the analyses presented in this work are publicly available. 
The data analysis pipelines and notebooks are available at \href{https://github.com/robertonetti/evo-intermediate-reconstruction}{\texttt{evo-intermediate-reconstruction}}.
The modified arDCA framework adapted for conditional inference of evolutionary intermediate sequences is available at \href{https://github.com/robertonetti/arDCA-evo-intermediate-reconstruction}{\texttt{arDCA-evo-intermediate-reconstruction}}. 
The datasets used in this study are deposited on Zenodo: \href{https://doi.org/10.5281/zenodo.20730522}{\texttt{data-evo-intermediate-reconstruction}}.

\medskip
\noindent
{\bf Acknowledgments --}
We thank Francesco Calvanese, Matteo Bisardi, Leonardo Di Bari, Lorenzo Rosset, Francesco Zamponi for useful discussions and for assistance with the code for generative modeling and evolutionary simulations.
We acknowledge financial support from the Horizon Europe MSCA Staff Exchange project ``SIMBAD'' (grant agreement no.\ 101131463).

\medskip
\noindent
{\bf Author Contributions --}
Both authors contributed equally.

\medskip
\noindent
{\bf Conflicts of Interest --}
The authors declare no competing interests.

\bibliography{library_corrected}

\clearpage
\onecolumngrid

\renewcommand{\thesection}{S\arabic{section}}
\renewcommand{\thesubsection}{S\arabic{section}.\arabic{subsection}}
\renewcommand{\thesubsubsection}{S\arabic{section}.\arabic{subsection}.\arabic{subsubsection}}
\renewcommand{\thetable}{S\arabic{table}}
\renewcommand{\thefigure}{S\arabic{figure}}

\makeatletter
\renewcommand{\p@subsection}{}        
\renewcommand{\p@subsubsection}{}     
\makeatother
\setcounter{section}{0}
\setcounter{figure}{0}
\setcounter{table}{0}

\setcounter{section}{0}
\setcounter{figure}{0}
\setcounter{table}{0}

\section*{Supplementary Appendix}

\section{Protein families and multiple sequence alignments}
\label{sec:protein_families_msa}

\subsection{Chorismate Mutase}
Chorismate Mutases (CM; Pfam accession: PF01817) are enzymes that catalyze the conversion of chorismate to prephenate, a key step in the biosynthetic pathway of phenylalanine and tyrosine. This family is used as the main benchmark in our study. We used the Chorismate Mutase MSA previously analyzed in \cite{Russ2020, Netti2026HighEntropy}, containing 1,259 sequences.

\subsection{$\beta$-lactamase}
$\beta$-lactamases (Pfam accession: PF13354) are enzymes involved in bacterial resistance to $\beta$-lactam antibiotics. This Pfam family corresponds to the catalytic domain of class A $\beta$-lactamases. In this work, we used the same $\beta$-lactamase dataset previously analyzed in \cite{DiBari2024} containing 18,334 sequences.

\subsection{Response Regulator Receiver domain (RR)}
The Response Regulator Receiver Domain (RR; Pfam accession: PF00072) is a large protein domain family commonly found in two-component signal transduction systems. In this study, we focused on the shared RR domain. After removing duplicated sequences and sequences with more than 20\% gaps, the final MSA contains 49,528 sequences over $L = 112$ aligned residues.

\section{Training family-specific bmDCA models}
\label{sec:bmdca_inference}

For each protein family analyzed in this work, we inferred a family-specific statistical sequence landscape using Boltzmann Machine Direct Coupling Analysis (bmDCA). In particular, we used the fully connected implementation of the \texttt{adabmDCA 2.0} package \cite{Rosset2026adabmDCA2}. The resulting model defines a probability distribution over aligned amino-acid sequences,
\begin{equation}
P(S) = \frac{1}{Z}\exp[-\mathcal{H}(S)] ,
\end{equation}
where $Z$ is the partition function and $\mathcal{H}(S)$ is the Potts Hamiltonian
\begin{equation}
\mathcal{H}(S) =
-\sum_{i<j} J_{ij}(s_i,s_j)
-\sum_i h_i(s_i) .
\end{equation}
Here, $S=(s_1,\dots,s_L)$ denotes an aligned protein sequence, $h_i(s_i)$ are site-specific fields, and $J_{ij}(s_i,s_j)$ are pairwise couplings between residue positions. The fields account for the conservation patterns of individual sites, whereas the couplings capture pairwise statistical dependencies between positions, including epistatic constraints.
The bmDCA models were trained directly on the natural multiple sequence alignments described in Section~\ref{sec:protein_families_msa}. Model parameters were updated to match the empirical statistics of the natural MSA. Training was stopped using an early-stopping criterion based on the agreement between natural and model-generated two-point connected correlations. Specifically, for each model we computed the Pearson correlation between the connected two-point correlations measured in the natural MSA and those measured in samples generated from the bmDCA model. Training was stopped when this Pearson correlation reached 0.95.

The inferred bmDCA models play a central role throughout our framework. They define the artificial sequence landscapes in which evolutionary trajectories are simulated, with the statistical energy $\mathcal{H}(S)$ used as a proxy for negative fitness. These trajectories provide the ground-truth data used to train the intermediate reconstruction models. The same bmDCA landscapes are also used to score reconstructed midpoints: after predicting or sampling an intermediate sequence, we compute its statistical energy under the corresponding family-specific model and compare it with that of the simulated ground truth. This allows us to evaluate whether a predicted intermediate is not only close to the true midpoint in sequence identity, but also statistically plausible within the landscape that generated the trajectory.

\section{Generation of simulated evolutionary trajectories}
\label{sec:evolutionary_simulations}

Evolutionary trajectories were simulated in the bmDCA landscape inferred for each protein family, as described in Section~\ref{sec:bmdca_inference}, using the simulation framework introduced in \cite{DiBari2024}. The aim of this step was to generate controlled evolutionary paths in which the true intermediate sequence is known. This provides a ground-truth setting for testing how well an intermediate can be reconstructed from the two endpoints of a trajectory.

The evolutionary dynamics is defined at the nucleotide level, following \cite{DiBari2024}. Mutations are therefore proposed on codons rather than directly on amino acids, so that the constraints and degeneracy of the genetic code are explicitly included in the simulation. Single-nucleotide substitutions are sampled with a Gibbs step, while codon-level insertions and deletions are sampled with a Metropolis step. Insertions and deletions are handled in the aligned sequence representation: deletions introduce gaps, whereas insertions can fill existing gaps. This allows the simulation to include indel events while keeping all sequences in the same alignment frame.

The bmDCA energy is used to guide the evolutionary dynamics. As a result, the simulated trajectories explore the sequence landscape defined by the inferred Potts model: moves toward lower-energy regions are favored, whereas moves toward high-energy regions are accepted less often. The long-time equilibrium distribution of this dynamics is the bmDCA distribution \cite{DiBari2024},
\[
P(S) \propto \exp[-\mathcal{H}(S)] .
\]
Thus, the same model that was inferred from the natural MSA defines the artificial landscape in which evolution is simulated.

For each protein family, simulations were initialized from an equilibrium sample of the corresponding bmDCA model. This avoids starting all trajectories from a small number of natural sequences and avoids any out-of-equilibrium effect. Each Markov chain was then evolved for a large number of Monte Carlo steps, where one step corresponds to one attempted mutation, insertion, or deletion event.

\subsection{Construction of ground-truth trajectory triplets}
\label{subsec:trajectory_triplets}

From the simulated trajectories, we extracted triplets of sequences,
\[
(S_\mathrm{start}, S_\mathrm{mid}, S_\mathrm{end}) .
\]
The three sequences are ordered in time: $S_\mathrm{start}$ is the first sequence of the segment, $S_\mathrm{mid}$ is the temporal midpoint, and $S_\mathrm{end}$ is the final sequence. By construction, the same evolutionary time $\tau$ separates $S_\mathrm{start}$ from $S_\mathrm{mid}$ and $S_\mathrm{mid}$ from $S_\mathrm{end}$:
\[
S_\mathrm{start}
\;\xrightarrow{\;\tau\;}\;
S_\mathrm{mid}
\;\xrightarrow{\;\tau\;}\;
S_\mathrm{end}.
\]
Each triplet therefore defines a supervised intermediate-reconstruction problem: given the two endpoints, the task is to recover or sample the true temporal midpoint.

Triplets were extracted for several evolutionary timescales,
$\tau \in \{10^1,10^2,10^3,10^4,10^5,10^6\}$
Monte Carlo steps. These values cover a broad range of dynamical regimes. At short times, the three sequences are close to each other and the reconstruction problem is mostly local. At intermediate times, the trajectory explores larger portions of the landscape. At long times, the endpoints and the midpoint become weakly correlated, so that reconstructing the true midpoint from the endpoints becomes intrinsically difficult.

For each value of $\tau$, we constructed a separate dataset of trajectory triplets, which was divided into training, validation, and test sets. The training set was used to learn the conditional autoregressive model for intermediate reconstruction, the validation set was used to monitor training, and the test set was kept separate for all reported reconstruction results.

This construction gives direct access to the true intermediate sequence $S_\mathrm{mid}$ for every pair of endpoints. It therefore allows us to measure reconstruction accuracy in a controlled setting and to compare the statistical properties of predicted intermediates with those of the simulated ground truth. In particular, we evaluate predicted sequences using their Hamming distance from the endpoints and their statistical energy under the same bmDCA landscape that generated the trajectories.

\section{Autoregressive modeling of evolutionary intermediates}
\label{sec:ardca_intermediate_reconstruction}

To reconstruct evolutionary intermediates from trajectory endpoints, we trained autoregressive DCA models (arDCA) \cite{Trinquier2021}. The training data consist of the simulated trajectory triplets described in Section~\ref{subsec:trajectory_triplets}, of the form
\[
(S_\mathrm{start},S_\mathrm{mid},S_\mathrm{end}) .
\]
Since the goal is to predict the temporal midpoint from the two endpoints, each triplet was reordered as
\[
(S_\mathrm{start},S_\mathrm{end},S_\mathrm{mid}) .
\]
With this ordering, the autoregressive model first reads the starting sequence, then the endpoint sequence, and finally learns to generate the midpoint sequence conditioned on both boundaries.

The model therefore represents the joint probability of the concatenated sequence as an autoregressive product. The part used for reconstruction is the conditional distribution
\[
P(S_\mathrm{mid}\mid S_\mathrm{start},S_\mathrm{end}) .
\]
This distribution is factorized site by site as
\[
P(S_\mathrm{mid}\mid S_\mathrm{start},S_\mathrm{end})
=
\prod_{i=1}^{L}
P(s^\mathrm{mid}_i
\mid
S_\mathrm{start},S_\mathrm{end},S^\mathrm{mid}_{<i}) .
\]
Thus, each residue of the midpoint is predicted using the two full endpoint sequences and the residues of the midpoint that have already been generated.

In practice, we modified the standard arDCA training objective so that the likelihood is computed only on the final block of the concatenated sequence, corresponding to $S_\mathrm{mid}$. This directly optimizes the conditional probability of the midpoint given the two endpoints, rather than the likelihood of the full triplet. This choice is also computationally more efficient, because the model is trained only on the part of the sequence that needs to be reconstructed.

\subsection{Training procedure for conditional arDCA models}
\label{subsec:ardca_training_details}

A separate arDCA model was trained for each protein family and each evolutionary timescale. Thus, for a given family, independent models were trained for
\[
\tau \in \{10^1,10^2,10^3,10^4,10^5,10^6\}.
\]
Each model was trained on the corresponding training set of simulated triplets described in Section~\ref{subsec:trajectory_triplets}, validated on the corresponding validation set, and evaluated only on the held-out test set.

We tested several values of the regularization parameters for all protein families and timescales. In the final experiments, we used an $L_2$ regularization of $10^{-3}$ on the couplings $J$ and $10^{-5}$ on the fields $h$. These values gave stable training and good validation performance across all families and all evolutionary timescales. Further details on the arDCA regularization scheme are given in \cite{Trinquier2021}.

Training was stopped using early stopping on the validation set. In all cases, we used a patience of 5 validation checks: training was stopped when the validation loss did not improve for 5 consecutive checks. The model with the best validation performance was then retained and used for reconstruction on the test set.

After training, each arDCA model was used in two ways. First, we performed greedy maximum-likelihood reconstruction, where each residue of $S_\mathrm{mid}$ is chosen as the most likely amino acid conditioned on the endpoints and on the previously reconstructed residues. Second, we sampled from the learned conditional distribution at temperature $T=1$, generating one stochastic midpoint for each pair of endpoints. The first procedure gives a deterministic point prediction, whereas the second generates typical samples from the posterior distribution learned by the model.

\subsection{Single-endpoint conditioning for ablation analyses}
\label{subsec:one_endpoint_models}

For the endpoint-ablation analysis, we also trained models conditioned on a single boundary sequence. The aim of this analysis was to quantify how much information about the midpoint is contained in one endpoint alone, and to compare this with the reconstruction obtained when both endpoints are provided.

In practice, the one-endpoint models were trained only in one temporal direction, using pairs ordered as
\[
(S_\mathrm{start},S_\mathrm{mid}) .
\]
With this ordering, the model learns the conditional distribution
\[
P(S_\mathrm{mid}\mid S_\mathrm{start}) .
\]
This is sufficient for our purpose because the evolutionary dynamics used to generate the trajectories is invariant under time reversal \cite{DiBari2024}. As a consequence, training on pairs ordered as $(S_\mathrm{start},S_\mathrm{mid})$ is equivalent to training on pairs ordered as $(S_\mathrm{end},S_\mathrm{mid})$, since the two sequences play interchangeable roles under the reversed dynamics. In other words, the same one-endpoint model can be used to evaluate the information contained in either boundary sequence.

At test time, we applied this model to the same held-out triplets used in the two-endpoint reconstruction experiments. For the start-only condition, the model was conditioned on $S_\mathrm{start}$ and used to predict $S_\mathrm{mid}$. For the end-only condition, the same model was conditioned on $S_\mathrm{end}$ and used to predict $S_\mathrm{mid}$. The test sets were therefore identical to those used for the models conditioned on both endpoints, allowing a direct comparison between the three conditioning regimes: both endpoints, start only, and end only.

The same training procedure, regularization parameters, and early-stopping criterion described in Section~\ref{subsec:ardca_training_details} were used for the one-endpoint models as for the two-endpoint models. Comparing their reconstruction accuracies allowed us to identify which endpoint carries more information about the true intermediate sequence in different regions of the landscape.

\section{Context-dependent entropy and landscape topology}
\label{sec:cde_landscape_topology}

\subsection{Definition of CDE}
\label{subsec:cde_definition}

We use the average context-dependent entropy, $\overline{\mathrm{CDE}}$, to characterize the local mutability of a sequence in the bmDCA landscape \cite{Figliuzzi2016, Vigue2022, RodriguezRivas2022, Poelwijk2016}. For a given sequence $S=(s_1,\dots,s_L)$, the CDE of site $i$ is defined as the entropy of the conditional distribution of the residue at that site, given the rest of the sequence:
\[
\mathrm{CDE}_i(S)
=
-\sum_{a}
P(s_i=a \mid S_{\setminus i})
\log P(s_i=a \mid S_{\setminus i}) .
\]
Here, $S_{\setminus i}$ denotes the sequence context obtained by removing site $i$, and the sum runs over all amino-acid states. The conditional probabilities are computed from the family-specific bmDCA model described in Section~\ref{sec:bmdca_inference}. In practice, for each site $i$, the model assigns a probability to each possible amino acid given the residues present at all other positions.

The CDE therefore measures how constrained a site is in its current sequence background. A low value means that only one or a few amino acids are compatible with the surrounding sequence context. A high value means that several amino acids remain possible at that position. Thus, CDE provides a local estimate of how many mutations are tolerated by the current sequence background \cite{Figliuzzi2016, Poelwijk2016}.

For each sequence, we average the site-wise CDE values over all positions:
\[
\overline{\mathrm{CDE}}(S)
=
\frac{1}{L}\sum_{i=1}^{L}\mathrm{CDE}_i(S) .
\]
This gives a single sequence-level measure of mutability. Sequences with low $\overline{\mathrm{CDE}}$ are more constrained and lie in basin-like regions of the landscape. Sequences with high $\overline{\mathrm{CDE}}$ are more permissive and lie in plateau-like regions, where more mutations are compatible with the sequence context \cite{deVisserKrug2014, DiBari2024, Rossi2025}.

\subsection{Relationship between context-dependent entropy and statistical energy}
\label{subsec:cde_energy_relationship}

\begin{figure}[h]
    \centering
    \includegraphics[width=1\linewidth]{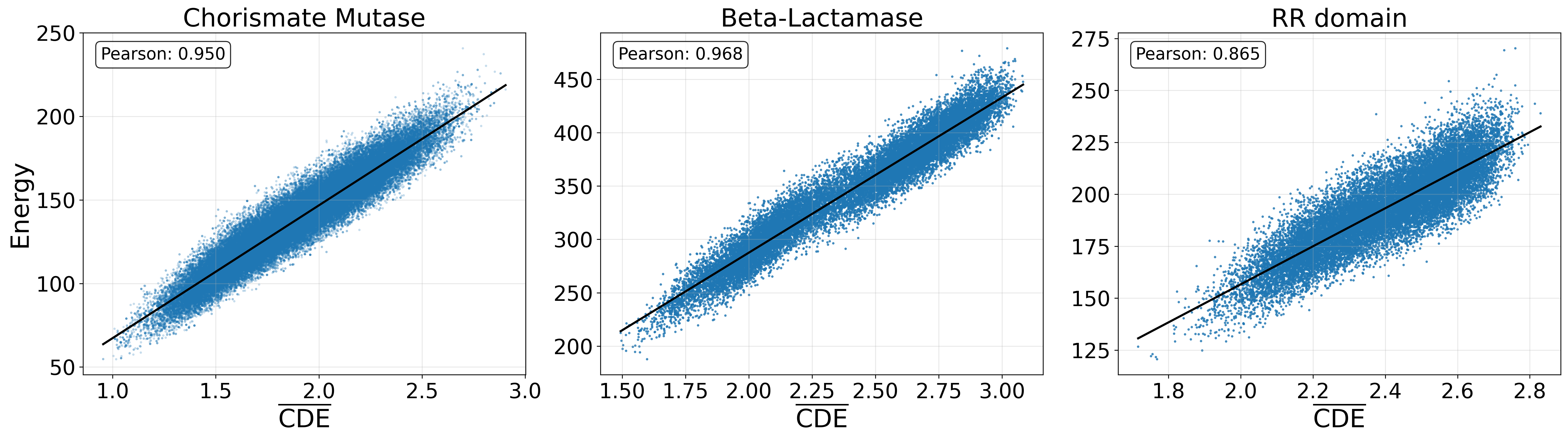}
    \caption{\textbf{Relationship between average CDE and bmDCA statistical energy.}
    Each subplot shows an equilibrium sample from one family-specific bmDCA model. For each sequence, the $x$-axis reports the average context-dependent entropy, $\overline{\mathrm{CDE}}$, and the $y$-axis reports the bmDCA statistical energy. The strong correlation observed in all three families shows that low-energy sequences tend to have low $\overline{\mathrm{CDE}}$, while higher-energy accessible sequences tend to have higher $\overline{\mathrm{CDE}}$.}
    \label{fig:cde_energy_all_families}
\end{figure}

The average CDE is closely related to the statistical energy assigned by the bmDCA model. To quantify this relationship, we computed $\overline{\mathrm{CDE}}$ and bmDCA energy for equilibrium samples from each family-specific model. The results are shown in Fig.~\ref{fig:cde_energy_all_families}.

Across the three protein families, sequences with lower statistical energy tend to have lower $\overline{\mathrm{CDE}}$, while sequences with higher statistical energy tend to have higher $\overline{\mathrm{CDE}}$. The correlation is strong in all cases. This supports the landscape interpretation used in the main text: low-energy regions correspond to more constrained basins, whereas higher-energy but still accessible regions correspond to more permissive plateaus \cite{deVisserKrug2014, DiBari2024, Rossi2025}.

This relationship is important for the interpretation of the evolutionary dynamics. In low-CDE regions, the local sequence context restricts the number of accessible mutations, so trajectories tend to move more slowly and retain more information about their past. In high-CDE regions, more residue states are compatible with the current context, so many alternative routes become possible. These regions are therefore more mutationally permissive and can weaken the information carried by the endpoints about the specific path followed by the trajectory \cite{DiBari2024, Rossi2025}.

\subsection{Definition of constrained and permissive CDE regimes}
\label{subsec:low_high_cde_regimes}

To classify regions of the landscape, we used the distribution of $\overline{\mathrm{CDE}}$ in an equilibrium sample of each family-specific bmDCA model. The use of $\overline{\mathrm{CDE}}$ as a landscape descriptor is motivated by its relationship with statistical energy, shown in Section~\ref{subsec:cde_energy_relationship}, and by its interpretation as a measure of sequence-context-dependent mutability \cite{Figliuzzi2016, Poelwijk2016, Vigue2022, RodriguezRivas2022}. This gives a family-specific reference distribution for sequence mutability. We then defined low- and high-CDE regimes using percentile thresholds of this distribution.

For each family, sequences below the 40th percentile of the equilibrium $\overline{\mathrm{CDE}}$ distribution were classified as low-CDE sequences. These sequences define the constrained regime, or basin-like part of the accessible landscape. Sequences above the 60th percentile were classified as high-CDE sequences. These sequences define the more permissive plateau-like regime. Sequences between the 40th and 60th percentiles were not assigned to either class in the basin--plateau analyses.

The gap between the two thresholds avoids classifying small fluctuations around the center of the distribution as transitions between regimes. This makes the classification more conservative: a trajectory is counted as moving from a basin to a plateau only when the endpoint sequences lie clearly on opposite sides of the equilibrium CDE distribution.

\begin{figure}[h]
    \centering
    \includegraphics[width=1\linewidth]{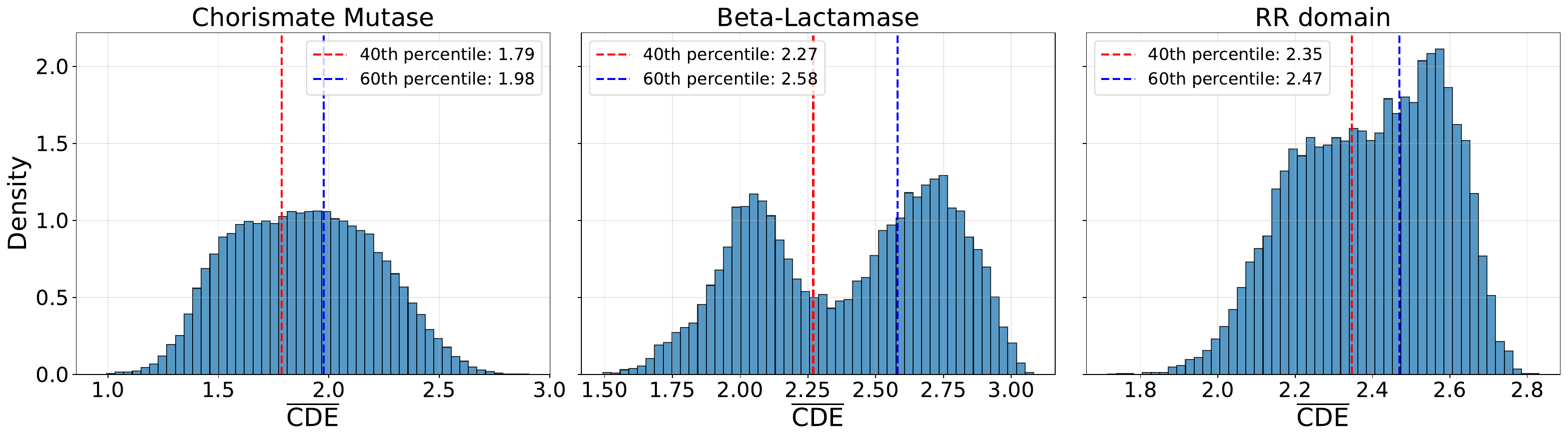}
    \caption{\textbf{Definition of low- and high-CDE regimes.}
    Each subplot shows the distribution of average context-dependent entropy, $\overline{\mathrm{CDE}}$, for an equilibrium sample from one family-specific bmDCA model. The two vertical lines indicate the 40th and 60th percentiles of the distribution. Sequences below the 40th percentile are classified as low-CDE, basin-like sequences, whereas sequences above the 60th percentile are classified as high-CDE, plateau-like sequences.}
    \label{fig:cde_distribution_all_families}
\end{figure}

The threshold definitions are shown in Fig.~\ref{fig:cde_distribution_all_families}. Each panel reports the equilibrium distribution of $\overline{\mathrm{CDE}}$ for one protein family, together with the 40th and 60th percentile cutoffs used to define the low- and high-CDE regimes.

\section{Benchmarking intermediate reconstruction strategies}
\label{sec:benchmark_reconstruction}

In the main text, we compared three strategies for reconstructing evolutionary intermediates in Chorismate Mutase: the direct-path baseline, greedy maximum-likelihood arDCA reconstruction, and generative sampling from the conditional arDCA model. Here, we report the same benchmark for the two additional protein families, $\beta$-lactamase and RR domain.

The goal of this comparison is to check whether the behavior observed in the main-text example is specific to Chorismate Mutase or is also recovered in other inferred landscapes. For each family, we evaluate the same quantities used in Fig.~2 of the main text: reconstruction accuracy, Hamming distance from the starting sequence, statistical energy under the corresponding ground-truth bmDCA model, and average pairwise divergence among reconstructed sequences.

Across both additional families, the same qualitative pattern is observed. The direct-path baseline performs well only at short timescales, where the endpoints are close and most differences can be interpreted as single substitutions. At longer timescales, it becomes less realistic because it ignores recurrent mutations and epistatic constraints. Greedy arDCA gives high residue-level accuracy, but tends to produce atypical intermediates, especially in terms of statistical energy and sequence diversity. In contrast, arDCA sampling better reproduces the global statistical properties of the simulated trajectories, even when its pointwise accuracy is slightly lower than that of greedy decoding.

\subsection{Reconstruction benchmark for $\beta$-lactamase}

\begin{figure}[h!]
    \centering
    \includegraphics[width=1\linewidth]{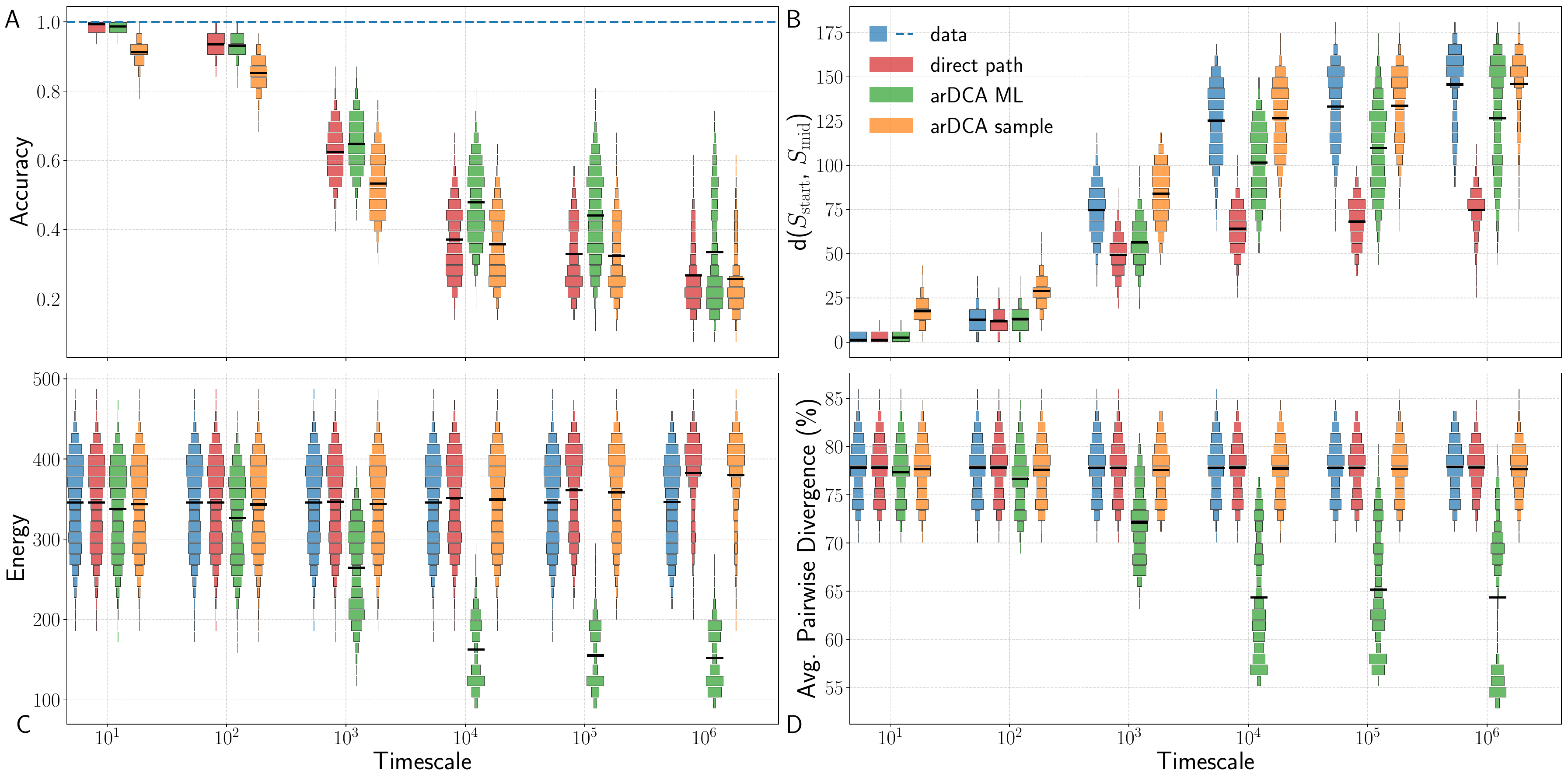}
    \caption{\textbf{Benchmark of reconstruction strategies in $\beta$-lactamase.}
    Comparison of direct-path reconstruction, greedy maximum-likelihood arDCA reconstruction, and generative sampling from the conditional arDCA model. The same metrics as in Fig.~2 of the main text are shown: reconstruction accuracy, Hamming distance from $S_\mathrm{start}$, statistical energy under the ground-truth bmDCA model, and average pairwise divergence among reconstructed sequences.}
    \label{fig:betalactamase_methods_comparison}
\end{figure}

Figure~\ref{fig:betalactamase_methods_comparison} shows the full comparison of reconstruction strategies for $\beta$-lactamase. The same four metrics used in the main text are reported. The results are consistent with the Chorismate Mutase analysis: direct paths become less reliable as the evolutionary timescale increases, while greedy arDCA and arDCA sampling differ in the tradeoff between pointwise accuracy and statistical realism.

\subsection{Reconstruction benchmark for RR domain}

\begin{figure}[h!]
    \centering
    \includegraphics[width=1\linewidth]{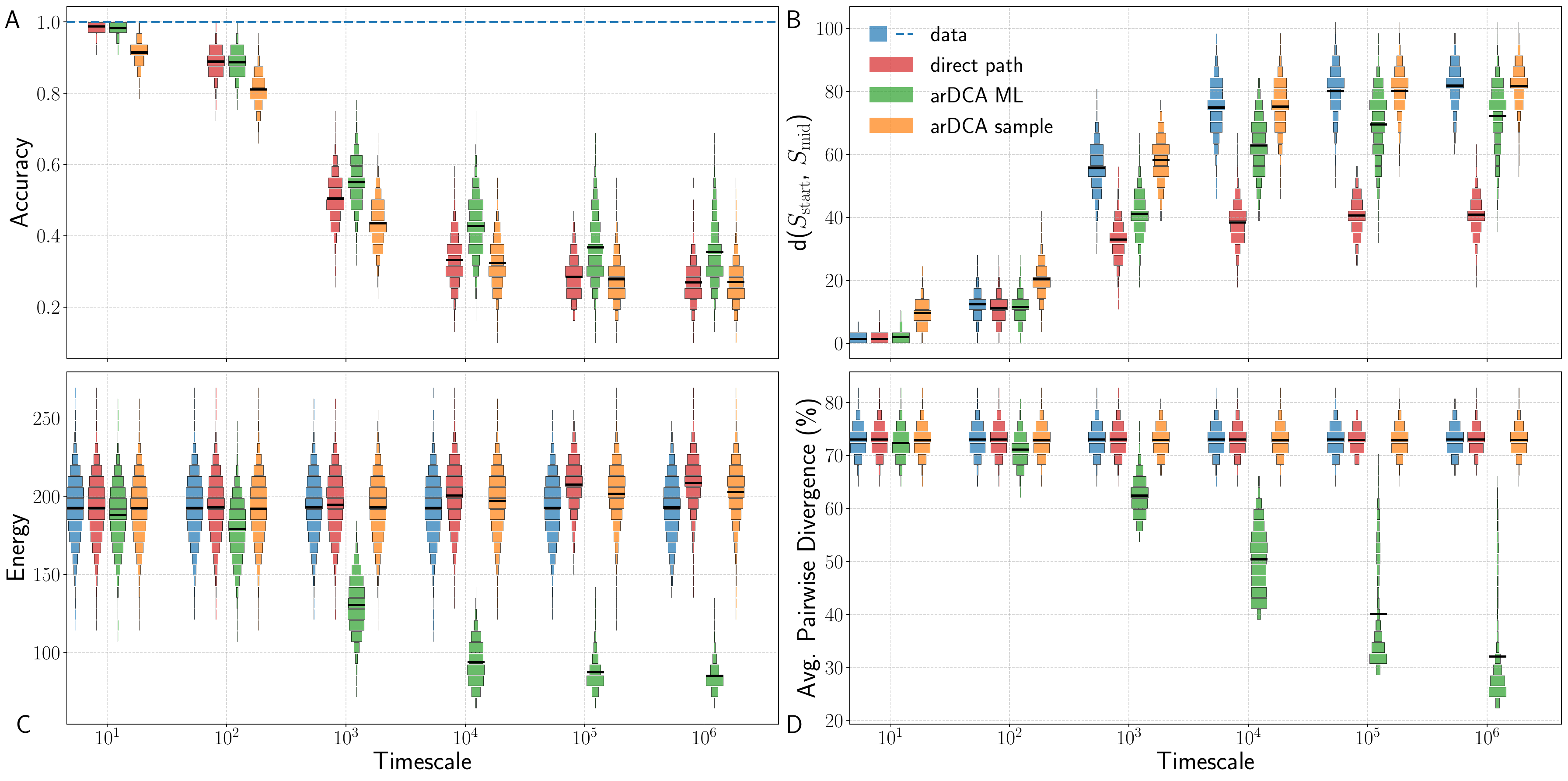}
    \caption{\textbf{Benchmark of reconstruction strategies in RR domain.}
    Comparison of direct-path reconstruction, greedy maximum-likelihood arDCA reconstruction, and generative sampling from the conditional arDCA model. The same metrics as in Fig.~2 of the main text are shown: reconstruction accuracy, Hamming distance from $S_\mathrm{start}$, statistical energy under the ground-truth bmDCA model, and average pairwise divergence among reconstructed sequences.}
    \label{fig:pf00072_methods_comparison}
\end{figure}

Figure~\ref{fig:pf00072_methods_comparison} reports the same benchmark for RR domain. Also in this family, the three reconstruction strategies show the same qualitative differences observed in the main text. This supports the generality of the conclusions drawn from Chorismate Mutase: greedy decoding is best suited for maximizing residue-level accuracy, whereas posterior sampling gives a more realistic reconstruction of the ensemble properties of evolutionary intermediates.

\newpage

\section{Endpoint ablation analysis and information asymmetry}
\label{sec:endpoint_ablation}

\subsection{Chorismate Mutase}

\subsubsection{Endpoint ablation on the complete Chorismate Mutase test set}

Before focusing on specific basin--plateau transitions, we first evaluated endpoint ablation on the complete unfiltered test set of Chorismate Mutase. In this setting, all trajectory triplets are included, independently of their $\overline{\mathrm{CDE}}$ values. This provides a baseline comparison for the three conditioning regimes: both endpoints, start only, and end only.

As shown in Table~\ref{tab:cm_ablation_unfiltered}, conditioning on both endpoints always gives the highest reconstruction accuracy across all evolutionary timescales. This confirms that, when no topological filtering is applied, the two boundary sequences provide complementary information about the temporal midpoint. The information asymmetry discussed in the main text therefore emerges only after separating trajectories according to the landscape regimes visited by the endpoints.

\begin{table}[h]
\caption{\textbf{Chorismate Mutase: endpoint ablation on the complete unfiltered test set.}
Reconstruction accuracy of autoregressive models conditioned on both endpoints, on $S_\mathrm{start}$ only, or on $S_\mathrm{end}$ only. All test triplets are included, without filtering by $\overline{\mathrm{CDE}}$. Conditioning on both endpoints gives the best reconstruction accuracy at all timescales.}
\label{tab:cm_ablation_unfiltered}

\begin{tabular*}{\columnwidth}{@{\extracolsep\fill}lcccccc@{}}
\toprule
\textbf{Conditioning} &
\multicolumn{6}{c}{\textbf{Evolutionary timescale ($\tau$)}} \\
\cmidrule(lr){2-7}
& $10^1$ & $10^2$ & $10^3$ & $10^4$ & $10^5$ & $10^6$ \\
\midrule
Both endpoints & \textbf{0.9843} & \textbf{0.9025} & \textbf{0.6451} & \textbf{0.5259} & \textbf{0.4232} & \textbf{0.2861} \\
Start only     & 0.9787 & 0.8898 & 0.6161 & 0.5123 & 0.3893 & 0.2854 \\
End only       & 0.9785 & 0.8861 & 0.6140 & 0.5120 & 0.3908 & 0.2850 \\
\bottomrule
\end{tabular*}
\end{table}

\subsubsection{Basin-to-plateau and plateau-to-basin transitions in Chorismate Mutase}

In the main text, we report the reconstruction accuracies obtained after filtering the test set into basin-to-plateau and plateau-to-basin transitions. Here, we report the number of test triplets available in each filtered class. These counts are important because some transition classes are absent, or represented by too few examples, at short evolutionary timescales. In such cases, the corresponding accuracy values are not meaningful and were therefore omitted from the main-text table.

For basin-to-plateau transitions, defined as low $\overline{\mathrm{CDE}} \to$ high $\overline{\mathrm{CDE}}$, no triplet is present at $\tau=10^1$, and only one triplet is present at $\tau=10^2$ (Table~\ref{tab:cm_basin_plateau_counts}). For plateau-to-basin transitions, defined as high $\overline{\mathrm{CDE}} \to$ low $\overline{\mathrm{CDE}}$, no test triplets are present at $\tau=10^1$ or $\tau=10^2$. For this reason, the corresponding short-timescale values are not used in the main-text accuracy table. From $\tau=10^3$ onward, the number of test triplets becomes large enough to provide meaningful reconstruction statistics in both transition classes.

\begin{table}[h]
\caption{\textbf{Chorismate Mutase: number of test triplets in basin--plateau transitions.}
Number of filtered test triplets used for the low $\overline{\mathrm{CDE}} \to$ high $\overline{\mathrm{CDE}}$ and high $\overline{\mathrm{CDE}} \to$ low $\overline{\mathrm{CDE}}$ analyses. The reconstruction accuracies for these same filtered datasets are reported in the main text.}
\label{tab:cm_basin_plateau_counts}

\begin{tabular*}{\columnwidth}{@{\extracolsep\fill}lcccccc@{}}
\toprule
\textbf{Transition class} &
\multicolumn{6}{c}{\textbf{Evolutionary timescale ($\tau$)}} \\
\cmidrule(lr){2-7}
& $10^1$ & $10^2$ & $10^3$ & $10^4$ & $10^5$ & $10^6$ \\
\midrule
Basin $\to$ Plateau &
0 & 1 & 143 & 1074 & 2636 & 3004 \\
Plateau $\to$ Basin &
0 & 0 & 122 & 1008 & 2489 & 3181 \\
\bottomrule
\end{tabular*}
\end{table}

\subsection{$\beta$-lactamase}

We next repeated the same endpoint-ablation analysis on the $\beta$-lactamase family. This provides an additional test of whether the behavior observed for Chorismate Mutase is specific to that landscape, or whether the same qualitative trends are recovered in a different protein family.

\subsubsection{Endpoint ablation on the complete $\beta$-lactamase test set}

As a first step, we evaluated the three conditioning regimes on the complete unfiltered $\beta$-lactamase test set. This analysis includes all trajectory triplets, without separating them according to their $\overline{\mathrm{CDE}}$ values. It therefore serves as the direct analogue of the unfiltered Chorismate Mutase analysis.

The results are shown in Table~\ref{tab:betalactamase_ablation_unfiltered}. As in Chorismate Mutase, conditioning on both endpoints gives the highest reconstruction accuracy at every evolutionary timescale. This confirms that, at the level of the full unfiltered test set, the two boundary sequences provide complementary information about the temporal midpoint. The more asymmetric behavior discussed below therefore reflects the structure of specific landscape transitions, rather than a generic failure of one of the two endpoints to provide useful information.

\begin{table}[h]
\caption{\textbf{$\beta$-lactamase: endpoint ablation on the complete unfiltered test set.}
Reconstruction accuracy of autoregressive models conditioned on both endpoints, on $S_\mathrm{start}$ only, or on $S_\mathrm{end}$ only. All test triplets are included, without filtering by $\overline{\mathrm{CDE}}$. Conditioning on both endpoints gives the best reconstruction accuracy at all timescales.}
\label{tab:betalactamase_ablation_unfiltered}

\begin{tabular*}{\columnwidth}{@{\extracolsep\fill}lcccccc@{}}
\toprule
\textbf{Conditioning} &
\multicolumn{6}{c}{\textbf{Evolutionary timescale ($\tau$)}} \\
\cmidrule(lr){2-7}
& $10^1$ & $10^2$ & $10^3$ & $10^4$ & $10^5$ & $10^6$ \\
\midrule
Both endpoints & \textbf{0.9873} & \textbf{0.9310} & \textbf{0.6467} & \textbf{0.4787} & \textbf{0.4415} & \textbf{0.3355} \\
Start only     & 0.9810 & 0.9200 & 0.6071 & 0.4742 & 0.4257 & 0.3148 \\
End only       & 0.9810 & 0.9200 & 0.6066 & 0.4739 & 0.4261 & 0.3152 \\
\bottomrule
\end{tabular*}
\end{table}

\subsubsection{Endpoint ablation across basin--plateau transitions in $\beta$-lactamase}

We then applied the same CDE-based filtering used for Chorismate Mutase, separating trajectories into basin-to-plateau and plateau-to-basin transitions. The reconstruction accuracies for these filtered subsets are reported in Table~\ref{tab:low-highCDE-betalactamase}. Here, we first report the number of available test triplets in each class, since this determines which timescales can be interpreted reliably.

In $\beta$-lactamase, transitions between low- and high-$\overline{\mathrm{CDE}}$ regions are not observed in the test set at $\tau=10^1$, $\tau=10^2$, or $\tau=10^3$ (Table~\ref{tab:betalactamase_basin_plateau_counts}). This indicates that, for this family, the shortest simulated timescales are not sufficient for the trajectories to move between these two landscape regimes. The filtered analysis therefore starts at $\tau=10^4$, where both transition directions are represented by enough examples to compute meaningful reconstruction accuracies.

\begin{table}[h]
\caption{\textbf{$\beta$-lactamase: number of test triplets in basin--plateau transitions.}
Number of filtered test triplets used for the low $\overline{\mathrm{CDE}} \to$ high $\overline{\mathrm{CDE}}$ and high $\overline{\mathrm{CDE}} \to$ low $\overline{\mathrm{CDE}}$ analyses. The reconstruction accuracies for these same filtered datasets are reported in Table~\ref{tab:low-highCDE-betalactamase}.}
\label{tab:betalactamase_basin_plateau_counts}

\begin{tabular*}{\columnwidth}{@{\extracolsep\fill}lcccccc@{}}
\toprule
\textbf{Transition class} &
\multicolumn{6}{c}{\textbf{Evolutionary timescale ($\tau$)}} \\
\cmidrule(lr){2-7}
& $10^1$ & $10^2$ & $10^3$ & $10^4$ & $10^5$ & $10^6$ \\
\midrule
Basin $\to$ Plateau &
0 & 0 & 0 & 97 & 887 & 2506 \\
Plateau $\to$ Basin &
0 & 0 & 0 & 100 & 902 & 2470 \\
\bottomrule
\end{tabular*}
\end{table}

\begin{table}[h]
\caption{\textbf{$\beta$-lactamase: reconstruction accuracy under endpoint ablation.}
Reconstruction accuracy of autoregressive models conditioned on both endpoints, on $S_\mathrm{start}$ only, or on $S_\mathrm{end}$ only, for transitions between low- and high-$\overline{\mathrm{CDE}}$ regions of the landscape. Timescales $10^1$, $10^2$, and $10^3$ are omitted because no test triplets are available for these filtered transition classes.}
\label{tab:low-highCDE-betalactamase}

\begin{tabular*}{\columnwidth}{@{\extracolsep\fill}lccc@{\extracolsep\fill}}
\toprule
\textbf{Conditioning} & \multicolumn{3}{c}{\textbf{Evolutionary timescale ($\tau$)}} \\
\cmidrule(lr){2-4}
& $10^4$ & $10^5$ & $10^6$ \\
\midrule
\multicolumn{4}{@{}l}{\textbf{Basin $\to$ Plateau} \quad (Low $\overline{\mathrm{CDE}} \to$ High $\overline{\mathrm{CDE}}$)} \\
\addlinespace[2pt]
Both endpoints & \textbf{0.4840} & \textbf{0.4277} & \textbf{0.3273} \\
Start only     & \textbf{0.4859} & \textbf{0.4233} & \textbf{0.3200} \\
End only       & 0.4606 & 0.3304 & 0.2539 \\
\addlinespace[4pt]
\midrule
\multicolumn{4}{@{}l}{\textbf{Plateau $\to$ Basin} \quad (High $\overline{\mathrm{CDE}} \to$ Low $\overline{\mathrm{CDE}}$)} \\
\addlinespace[2pt]
Both endpoints & \textbf{0.4817} & \textbf{0.4368} & \textbf{0.3306} \\
Start only     & 0.4650 & 0.3375 & 0.2536 \\
End only       & \textbf{0.4838} & \textbf{0.4333} & \textbf{0.3241} \\
\bottomrule
\end{tabular*}

\end{table}

\subsection{RR domain}

We finally repeated the same endpoint-ablation analysis on the RR domain family. Together with Chorismate Mutase and $\beta$-lactamase, this provides a third test case for evaluating whether the information-asymmetry pattern is conserved across different inferred landscapes.

\subsubsection{Endpoint ablation on the complete RR domain test set}

We first evaluated the three conditioning regimes on the complete unfiltered RR domain test set. As in the previous families, this analysis includes all trajectory triplets, without separating them according to their $\overline{\mathrm{CDE}}$ values. It therefore provides the reference behavior before applying any topological filtering.

The results are shown in Table~\ref{tab:pf00072_ablation_unfiltered}. Conditioning on both endpoints gives the highest reconstruction accuracy from $\tau=10^1$ to $\tau=10^5$. At the longest timescale, $\tau=10^6$, the start-only and end-only models become marginally more accurate, but the differences are very small. This is expected in the near-decorrelated regime, where the endpoints contain little information about the true midpoint and the three conditioning regimes become almost equivalent.

\begin{table}[h]
\caption{\textbf{RR domain: endpoint ablation on the complete unfiltered test set.}
Reconstruction accuracy of autoregressive models conditioned on both endpoints, on $S_\mathrm{start}$ only, or on $S_\mathrm{end}$ only. All test triplets are included, without filtering by $\overline{\mathrm{CDE}}$. Conditioning on both endpoints gives the best reconstruction accuracy up to $\tau=10^5$, while the three conditioning regimes become nearly equivalent at $\tau=10^6$.}
\label{tab:pf00072_ablation_unfiltered}

\begin{tabular*}{\columnwidth}{@{\extracolsep\fill}lcccccc@{}}
\toprule
\textbf{Conditioning} &
\multicolumn{6}{c}{\textbf{Evolutionary timescale ($\tau$)}} \\
\cmidrule(lr){2-7}
& $10^1$ & $10^2$ & $10^3$ & $10^4$ & $10^5$ & $10^6$ \\
\midrule
Both endpoints & \textbf{0.9827} & \textbf{0.8871} & \textbf{0.5507} & \textbf{0.4279} & \textbf{0.3673} & 0.3552 \\
Start only     & 0.9775 & 0.8751 & 0.5193 & 0.4180 & 0.3658 & \textbf{0.3598} \\
End only       & 0.9774 & 0.8749 & 0.5196 & 0.4178 & 0.3652 & \textbf{0.3596} \\
\bottomrule
\end{tabular*}
\end{table}

\subsubsection{Endpoint ablation across basin--plateau transitions in RR domain}

We then applied the same CDE-based filtering to RR domain, separating trajectories into basin-to-plateau and plateau-to-basin transitions. The reconstruction accuracies for these filtered subsets are reported in Table~\ref{tab:low-highCDE-pf00072}. As above, we first report the number of available test triplets in each class, since the shortest timescales contain very few or no examples of transitions between low- and high-$\overline{\mathrm{CDE}}$ regions.

For basin-to-plateau transitions, defined as low $\overline{\mathrm{CDE}} \to$ high $\overline{\mathrm{CDE}}$, no test triplet is present at $\tau=10^1$, and only two triplets are present at $\tau=10^2$ (Table~\ref{tab:pf00072_basin_plateau_counts}). For plateau-to-basin transitions, defined as high $\overline{\mathrm{CDE}} \to$ low $\overline{\mathrm{CDE}}$, no test triplet is present at $\tau=10^1$, and only one triplet is present at $\tau=10^2$. These short-timescale values are therefore not interpreted in the filtered analysis. From $\tau=10^3$ onward, both transition directions are represented by enough examples to compute meaningful reconstruction accuracies.

\begin{table}[h]
\caption{\textbf{RR domain: number of test triplets in basin--plateau transitions.}
Number of filtered test triplets used for the low $\overline{\mathrm{CDE}} \to$ high $\overline{\mathrm{CDE}}$ and high $\overline{\mathrm{CDE}} \to$ low $\overline{\mathrm{CDE}}$ analyses. The reconstruction accuracies for these same filtered datasets are reported in Table~\ref{tab:low-highCDE-pf00072}.}
\label{tab:pf00072_basin_plateau_counts}

\begin{tabular*}{\columnwidth}{@{\extracolsep\fill}lcccccc@{}}
\toprule
\textbf{Transition class} &
\multicolumn{6}{c}{\textbf{Evolutionary timescale ($\tau$)}} \\
\cmidrule(lr){2-7}
& $10^1$ & $10^2$ & $10^3$ & $10^4$ & $10^5$ & $10^6$ \\
\midrule
Basin $\to$ Plateau &
0 & 2 & 271 & 1714 & 3172 & 3253 \\
Plateau $\to$ Basin &
0 & 1 & 303 & 1680 & 3225 & 3104 \\
\bottomrule
\end{tabular*}
\end{table}

\begin{table}[h!]
\caption{\textbf{RR domain: reconstruction accuracy under endpoint ablation.}
Reconstruction accuracy of autoregressive models conditioned on both endpoints, on $S_\mathrm{start}$ only, or on $S_\mathrm{end}$ only, for transitions between low- and high-$\overline{\mathrm{CDE}}$ regions of the landscape. Timescales $10^1$ and $10^2$ are omitted because no test triplets, or too few test triplets, are available for these filtered transition classes.}
\label{tab:low-highCDE-pf00072}

\begin{tabular*}{\columnwidth}{@{\extracolsep\fill}lcccc@{\extracolsep\fill}}
\toprule
\textbf{Conditioning} & \multicolumn{4}{c}{\textbf{Evolutionary timescale ($\tau$)}} \\
\cmidrule(lr){2-5}
& $10^3$ & $10^4$ & $10^5$ & $10^6$ \\
\midrule
\multicolumn{5}{@{}l}{\textbf{Basin $\to$ Plateau} \quad (Low $\overline{\mathrm{CDE}} \to$ High $\overline{\mathrm{CDE}}$)} \\
\addlinespace[2pt]
Both endpoints & \textbf{0.5474} & \textbf{0.4260} & \textbf{0.3688} & \textbf{0.3569} \\
Start only     & \textbf{0.5152} & \textbf{0.4206} & \textbf{0.3723} & \textbf{0.3625} \\
End only       & 0.5103 & 0.4028 & 0.3616 & 0.3601 \\
\addlinespace[4pt]
\midrule
\multicolumn{5}{@{}l}{\textbf{Plateau $\to$ Basin} \quad (High $\overline{\mathrm{CDE}} \to$ Low $\overline{\mathrm{CDE}}$)} \\
\addlinespace[2pt]
Both endpoints & \textbf{0.5390} & \textbf{0.4251} & \textbf{0.3703} & \textbf{0.3538} \\
Start only     & 0.5037 & 0.3998 & 0.3639 & 0.3556 \\
End only       & \textbf{0.5130} & \textbf{0.4186} & \textbf{0.3709} & \textbf{0.3585} \\
\bottomrule
\end{tabular*}

\end{table}

\section{Barrier-crossing trajectories and loss of path information}
\label{sec:barrier_crossing}

In the main text, we show that trajectories crossing a high-$\overline{\mathrm{CDE}}$ region are harder to reconstruct than trajectories that remain in constrained low-$\overline{\mathrm{CDE}}$ regions (see Section~\ref{subsec:low_high_cde_regimes}). One possible concern is that this effect could simply reflect larger endpoint divergence, since barrier-crossing trajectories may connect more distant endpoints.

To control for this, we compared constrained and barrier-crossing trajectories within the same Hamming-distance bins, where endpoint divergence is measured as $d(S_\mathrm{start},S_\mathrm{end})$. Constrained paths are defined as triplets in which $S_\mathrm{start}$, $S_\mathrm{mid}$, and $S_\mathrm{end}$ all belong to the low-$\overline{\mathrm{CDE}}$ regime. Barrier-crossing paths are defined as triplets in which the two endpoints are low-$\overline{\mathrm{CDE}}$ sequences, while $S_\mathrm{mid}$ lies in the high-$\overline{\mathrm{CDE}}$ regime.

\begin{figure}[h]
    \centering
    \includegraphics[width=1\linewidth]{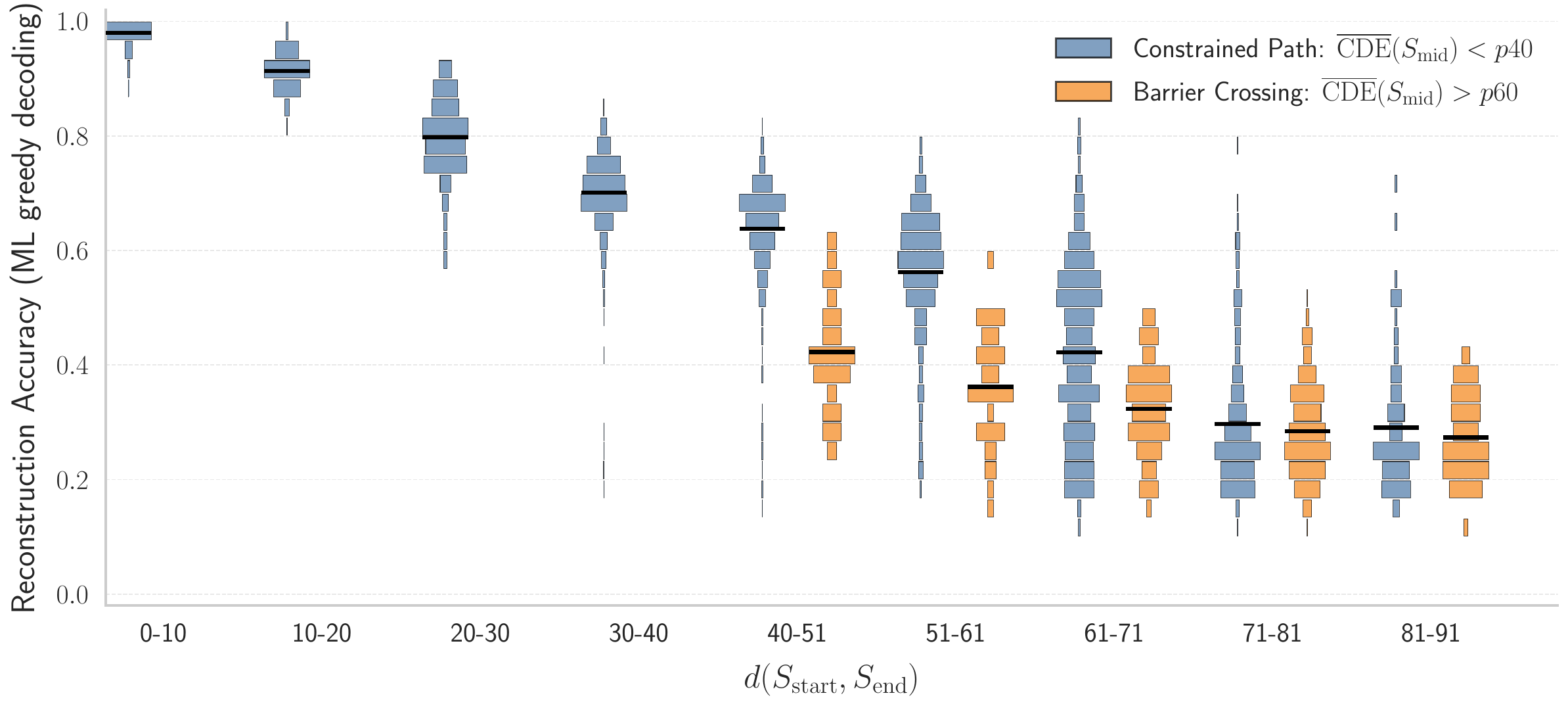}
    \caption{\textbf{Chorismate Mutase: barrier-crossing trajectories at fixed endpoint divergence.}
    Reconstruction accuracy is shown as a function of the Hamming distance between $S_\mathrm{start}$ and $S_\mathrm{end}$. Constrained paths remain in the low-$\overline{\mathrm{CDE}}$ regime at all three sampled points, whereas barrier-crossing paths have low-$\overline{\mathrm{CDE}}$ endpoints and a high-$\overline{\mathrm{CDE}}$ midpoint. Barrier-crossing paths remain harder to reconstruct within comparable distance bins.}
    \label{fig:cm_barrier_cde_accuracy_by_hamming_bins}
\end{figure}

\begin{figure}[h]
    \centering
    \includegraphics[width=1\linewidth]{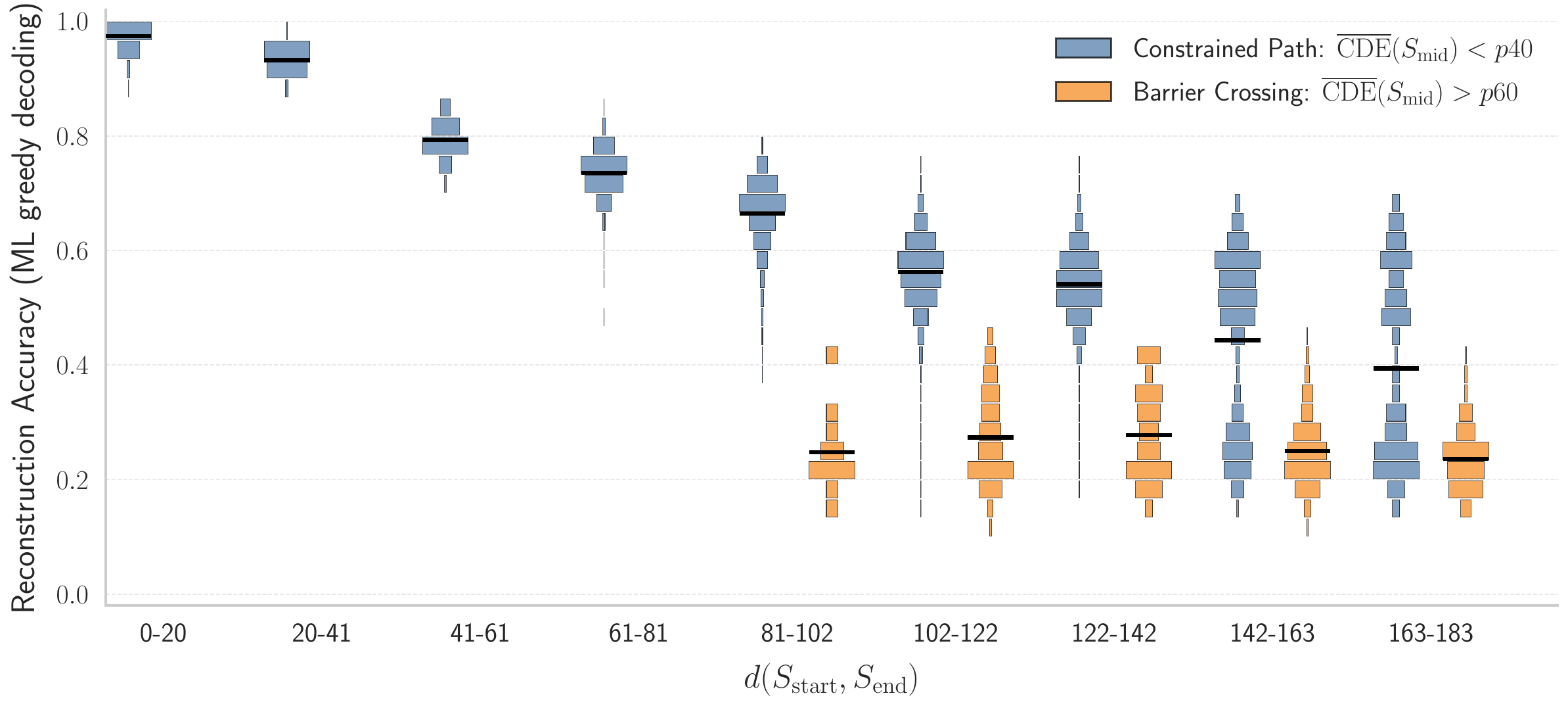}
    \caption{\textbf{$\beta$-lactamase: barrier-crossing trajectories at fixed endpoint divergence.}
    Reconstruction accuracy is shown as a function of the Hamming distance between $S_\mathrm{start}$ and $S_\mathrm{end}$. Barrier-crossing trajectories remain harder to reconstruct than constrained paths within comparable distance bins.}
    \label{fig:betalactamase_barrier_cde_accuracy_by_hamming_bins}
\end{figure}

Across the three families, barrier-crossing trajectories remain less accurately reconstructed than constrained trajectories at comparable endpoint divergence. This shows that the drop in accuracy is not explained by endpoint distance alone, but reflects a loss of path information associated with the high-$\overline{\mathrm{CDE}}$ excursion (Fig.~\ref{fig:cm_barrier_cde_accuracy_by_hamming_bins}, \ref{fig:betalactamase_barrier_cde_accuracy_by_hamming_bins}, \ref{fig:pf00072_barrier_cde_accuracy_by_hamming_bins}).

\subsection{Barrier-crossing analysis in Chorismate Mutase}

For Chorismate Mutase, barrier-crossing events are absent at $\tau=10^1$ and $\tau=10^2$, and only three such triplets are present at $\tau=10^3$ (Table~\ref{tab:cm_barrier_crossing_counts}). For this reason, the main-text comparison reports barrier-crossing accuracies from $\tau=10^4$ onward, where the number of events is sufficient for a reliable estimate.

\begin{table}[h]
\caption{\textbf{Chorismate Mutase: number of constrained and barrier-crossing triplets.}
The corresponding reconstruction accuracies are reported in the main text.}
\label{tab:cm_barrier_crossing_counts}

\begin{tabular*}{\columnwidth}{@{\extracolsep\fill}lcccccc@{}}
\toprule
\textbf{Trajectory class} &
\multicolumn{6}{c}{\textbf{Evolutionary timescale ($\tau$)}} \\
\cmidrule(lr){2-7}
& $10^1$ & $10^2$ & $10^3$ & $10^4$ & $10^5$ & $10^6$ \\
\midrule
Constrained path &
7913 & 7434 & 6506 & 4886 & 2740 & 1723 \\
Barrier crossing &
0 & 0 & 3 & 130 & 590 & 1077 \\
\bottomrule
\end{tabular*}
\end{table}

\subsection{Barrier-crossing analysis in $\beta$-lactamase}

In $\beta$-lactamase, barrier-crossing events are absent at $\tau=10^1$, $\tau=10^2$, and $\tau=10^3$ (Table~\ref{tab:betalactamase_barrier_crossing_counts}). From $\tau=10^4$ onward, these events become represented in the test set, allowing a direct comparison with constrained paths.

The reconstruction accuracies are reported in Table~\ref{tab:cde_barrier_accuracy_betalactamase}. The same pattern observed for Chorismate Mutase is recovered: barrier-crossing trajectories have lower reconstruction accuracy than constrained trajectories, despite both endpoints lying in low-$\overline{\mathrm{CDE}}$ regions.

\begin{table}[h]
\caption{\textbf{$\beta$-lactamase: number of constrained and barrier-crossing triplets.}}
\label{tab:betalactamase_barrier_crossing_counts}

\begin{tabular*}{\columnwidth}{@{\extracolsep\fill}lcccccc@{}}
\toprule
\textbf{Trajectory class} &
\multicolumn{6}{c}{\textbf{Evolutionary timescale ($\tau$)}} \\
\cmidrule(lr){2-7}
& $10^1$ & $10^2$ & $10^3$ & $10^4$ & $10^5$ & $10^6$ \\
\midrule
Constrained path &
3470 & 3362 & 3041 & 2596 & 2084 & 1498 \\
Barrier crossing &
0 & 0 & 0 & 48 & 338 & 1188 \\
\bottomrule
\end{tabular*}
\end{table}

\begin{table}[h]
\caption{\textbf{$\beta$-lactamase: reconstruction accuracy under barrier crossing.}
Reconstruction accuracy of greedy maximum-likelihood arDCA predictions for constrained and barrier-crossing paths. Barrier-crossing trajectories are absent at $\tau=10^1$, $\tau=10^2$, and $\tau=10^3$.}
\label{tab:cde_barrier_accuracy_betalactamase}

\begin{tabular*}{\columnwidth}{@{\extracolsep\fill}lcccccc@{}}
\toprule
\textbf{Trajectory class} &
\multicolumn{6}{c}{\textbf{Evolutionary timescale ($\tau$)}} \\
\cmidrule(lr){2-7}
& $10^1$ & $10^2$ & $10^3$ & $10^4$ & $10^5$ & $10^6$ \\
\midrule
Constrained path &
0.989 & 0.947 & 0.734 & 0.615 & 0.597 & 0.533 \\
Barrier crossing &
-- & -- & -- & 0.482 & 0.382 & 0.253 \\
\bottomrule
\end{tabular*}
\end{table}

\subsection{Barrier-crossing analysis in RR domain}

In RR domain, barrier-crossing events are absent at $\tau=10^1$ and $\tau=10^2$, while 30 examples are present at $\tau=10^3$ (Table~\ref{tab:pf00072_barrier_crossing_counts}). We report the corresponding accuracy for completeness, but this value should be interpreted with caution because of the small sample size.

\begin{figure}[h]
    \centering
    \includegraphics[width=1\linewidth]{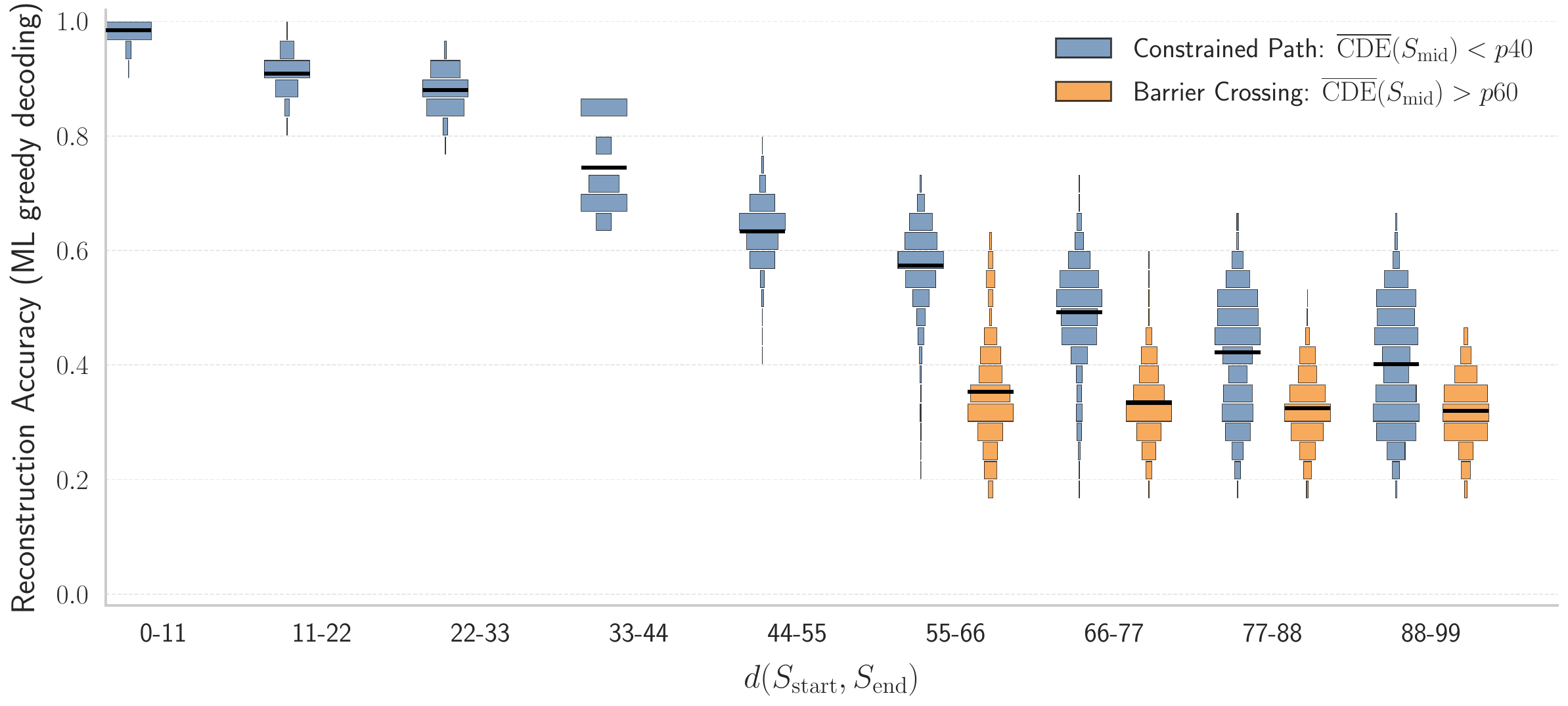}
    \caption{\textbf{RR domain: barrier-crossing trajectories at fixed endpoint divergence.}
    Reconstruction accuracy is shown as a function of the Hamming distance between $S_\mathrm{start}$ and $S_\mathrm{end}$. Barrier-crossing trajectories remain harder to reconstruct than constrained paths within comparable distance bins.}
    \label{fig:pf00072_barrier_cde_accuracy_by_hamming_bins}
\end{figure}

As shown in Table~\ref{tab:cde_barrier_accuracy_pf00072}, barrier-crossing trajectories are consistently less accurately reconstructed than constrained paths. The same trend is also visible when trajectories are compared at fixed endpoint divergence (Fig.~\ref{fig:pf00072_barrier_cde_accuracy_by_hamming_bins}).

\begin{table}[h]
\caption{\textbf{RR domain: number of constrained and barrier-crossing triplets.}}
\label{tab:pf00072_barrier_crossing_counts}

\begin{tabular*}{\columnwidth}{@{\extracolsep\fill}lcccccc@{}}
\toprule
\textbf{Trajectory class} &
\multicolumn{6}{c}{\textbf{Evolutionary timescale ($\tau$)}} \\
\cmidrule(lr){2-7}
& $10^1$ & $10^2$ & $10^3$ & $10^4$ & $10^5$ & $10^6$ \\
\midrule
Constrained path &
3096 & 2858 & 2295 & 1438 & 645 & 570 \\
Barrier crossing &
0 & 0 & 30 & 584 & 1627 & 1807 \\
\bottomrule
\end{tabular*}
\end{table}

\begin{table}[h]
\caption{\textbf{RR domain: reconstruction accuracy under barrier crossing.}
Reconstruction accuracy of greedy maximum-likelihood arDCA predictions for constrained and barrier-crossing paths. Barrier-crossing trajectories are absent at $\tau=10^1$ and $\tau=10^2$. At $\tau=10^3$, the estimate is based on 30 test triplets.}
\label{tab:cde_barrier_accuracy_pf00072}

\begin{tabular*}{\columnwidth}{@{\extracolsep\fill}lcccccc@{}}
\toprule
\textbf{Trajectory class} &
\multicolumn{6}{c}{\textbf{Evolutionary timescale ($\tau$)}} \\
\cmidrule(lr){2-7}
& $10^1$ & $10^2$ & $10^3$ & $10^4$ & $10^5$ & $10^6$ \\
\midrule
Constrained path &
0.985 & 0.906 & 0.624 & 0.529 & 0.456 & 0.417 \\
Barrier crossing &
-- & -- & 0.532 & 0.405 & 0.331 & 0.325 \\
\bottomrule
\end{tabular*}
\end{table}

\section{Distance-conditioned reconstruction of evolutionary intermediates}
\label{sec:distance_conditioned_reconstruction}

\subsection{Construction of distance-conditioned datasets}
\label{subsec:distance_conditioned_datasets}

In the main analysis, trajectory triplets are grouped by evolutionary time $\tau$. This is possible in simulations, because the elapsed Monte Carlo time between $S_\mathrm{start}$, $S_\mathrm{mid}$, and $S_\mathrm{end}$ is known by construction. In real applications, however, this time is not directly available. For this reason, we also built datasets in which triplets are grouped by endpoint divergence, measured as the Hamming distance between $S_\mathrm{start}$ and $S_\mathrm{end}$.

This analysis was performed for Chorismate Mutase. To obtain enough data in each divergence interval, we started from 200,000 equilibrium sequences sampled from the bmDCA model inferred for Chorismate Mutase. A larger initial sample was needed because the standard trajectory datasets were not sufficient to populate narrow Hamming-distance bins uniformly. Starting from these equilibrium sequences, we simulated evolutionary trajectories for $10^7$ Monte Carlo steps using the same nucleotide-level evolutionary dynamics described above and introduced in \cite{DiBari2024}.

From these trajectories, we extracted triplets
\[
(S_\mathrm{start},S_\mathrm{mid},S_\mathrm{end})
\]
at the same evolutionary timescales used in the time-conditioned analysis. The triplets from all timescales were then pooled together and reassigned to datasets according to the endpoint divergence $d(S_\mathrm{start},S_\mathrm{end})$. We used Hamming-distance intervals of width 5:
\[
[0,5), [5,10), \ldots, [65,70).
\]
The analysis was stopped at distance 70 because, beyond this value, the endpoint distance approaches the equilibrium saturation regime for Chorismate Mutase. In this regime, the same distance interval contains triplets coming from a broad range of evolutionary timescales, and the distance becomes much less informative about the underlying dynamics.

\begin{figure}[h!]
    \centering
    \includegraphics[width=1\linewidth]{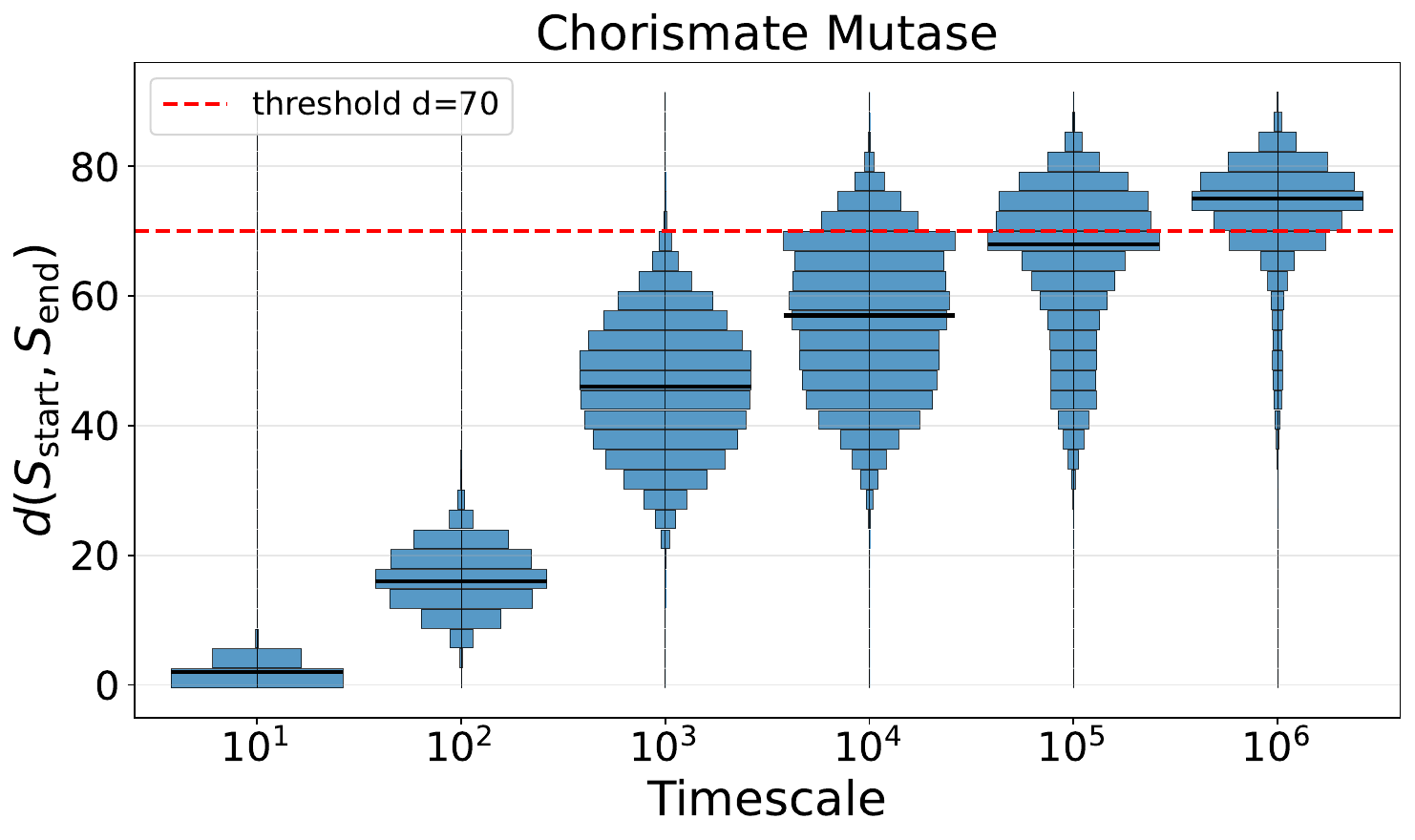}
    \caption{\textbf{Endpoint-divergence distributions across evolutionary timescales in Chorismate Mutase.}
    For each evolutionary timescale, the histogram shows the distribution of Hamming distances between $S_\mathrm{start}$ and $S_\mathrm{end}$. The red horizontal line indicates the cutoff at distance 70 used for the distance-conditioned reconstruction analysis. Beyond this cutoff, endpoint distances enter a saturation regime in which triplets from several evolutionary timescales, from $\tau=10^3$ to $\tau=10^6$, contribute to the same high-divergence range.}
    \label{fig:cm_distance_saturation_histograms}
\end{figure}

This effect is shown in Fig.~\ref{fig:cm_distance_saturation_histograms}. At small and intermediate endpoint distances, the distribution of distances still reflects the progression of evolutionary time. In contrast, for distances above approximately 70, triplets from several timescales, from $\tau=10^3$ to $\tau=10^6$, contribute to the same saturated distance regime. This motivates the cutoff used for the distance-conditioned reconstruction analysis.

The resulting distance-conditioned datasets were strongly unbalanced, because some divergence intervals contained many more triplets than others. To avoid training models on datasets with very different sizes, we balanced the bins by identifying the least populated distance interval and randomly subsampling the same number of triplets from all other intervals. After this balancing step, we constructed, for each endpoint-divergence bin, a training set of 3,500 triplets, a validation set of 500 triplets, and a test set of 1,500 triplets.

\subsection{Training of endpoint-divergence-conditioned arDCA models}
\label{subsec:distance_conditioned_ardca}

For each Hamming-distance bin, we trained a separate arDCA model to reconstruct $S_\mathrm{mid}$ from $S_\mathrm{start}$ and $S_\mathrm{end}$. The input ordering was the same as in the time-conditioned reconstruction experiments:
\[
(S_\mathrm{start},S_\mathrm{end},S_\mathrm{mid}) .
\]
The only difference is the criterion used to define the training dataset. In the time-conditioned analysis, each model is trained on triplets belonging to a fixed evolutionary timescale $\tau$. In the distance-conditioned analysis, each model is trained on triplets belonging to a fixed endpoint-divergence interval.

All distance-conditioned models were trained using the same training settings described in Section~\ref{subsec:ardca_training_details} for the time-conditioned arDCA models. In particular, the same regularization parameters, validation procedure, and early-stopping criterion were used. Thus, the comparison between time-conditioned and distance-conditioned reconstruction does not depend on different training settings.

\subsection{Statistical properties of distance-conditioned reconstructions}
\label{subsec:distance_conditioned_statistics}

We then compared the statistical properties of intermediates generated by distance-conditioned models with those generated by time-conditioned models. For the latter, the evolutionary time was not taken from the simulation labels, but inferred from endpoint features. This timescale was predicted from endpoint observables using the multinomial logistic regression described in Section~\ref{sec:timescale_inference}. In particular, the classifier uses the binary mismatch vector between the two endpoints, together with the residue-level CDE profiles of $S_\mathrm{start}$ and $S_\mathrm{end}$, as defined in Section~\ref{sec:timescale_feature_sets}.

The comparison was performed on the held-out test sets of the distance-conditioned datasets. For each test triplet, we reconstructed $S_\mathrm{mid}$ using either the model trained directly on the corresponding distance bin, or the time-conditioned model selected from the inferred timescale. We then compared the predicted intermediates with the simulated ground truth using the same quantities used in the main analysis: Hamming distance from $S_\mathrm{start}$ and statistical energy under the bmDCA landscape.

\begin{figure}[h]
    \centering
    \includegraphics[width=1\linewidth]{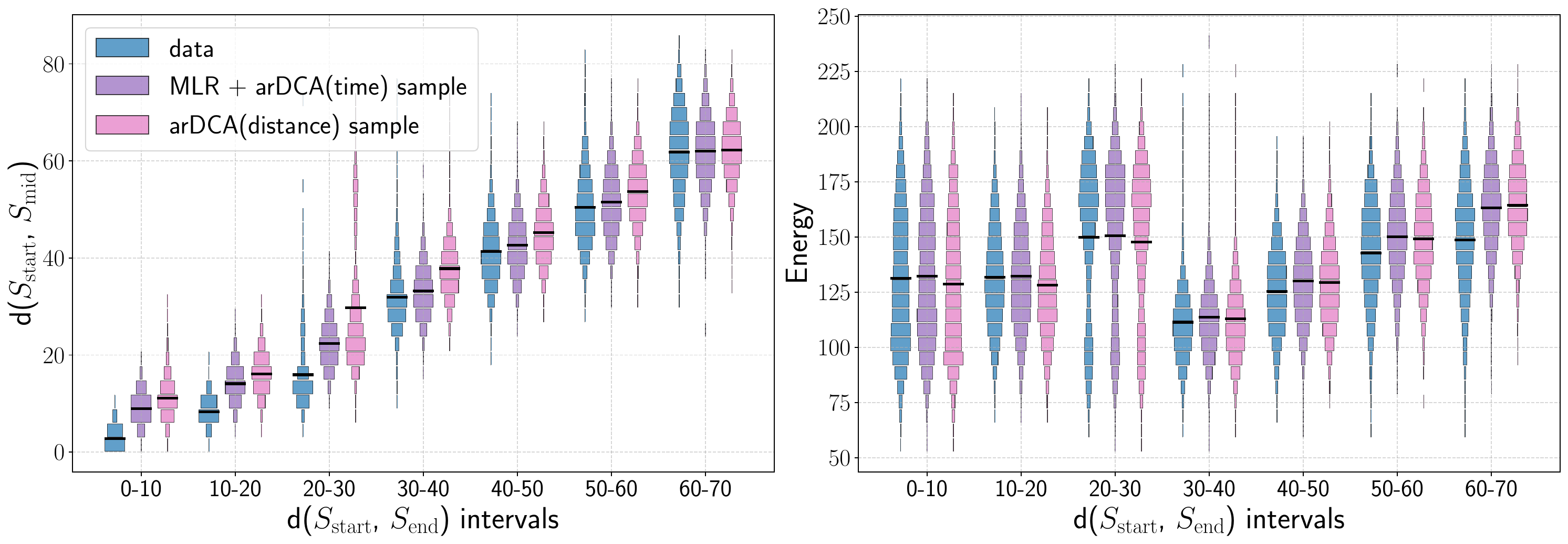}
    \caption{\textbf{Comparison of time- and distance-conditioned reconstructions.}
    Comparison between simulated ground-truth intermediates, reconstructions obtained with the time-conditioned model selected by multinomial logistic regression, MLR + arDCA(time), and reconstructions obtained with models trained directly on endpoint-divergence bins, arDCA(distance). The $x$-axis shows bins of Hamming divergence between $S_\mathrm{start}$ and $S_\mathrm{end}$. Left: Hamming distance between $S_\mathrm{start}$ and $S_\mathrm{mid}$ for the ground truth and for the two reconstruction strategies. Right: statistical energy of the reconstructed or simulated midpoint, computed with the ground-truth bmDCA model. For readability, adjacent endpoint-divergence bins are grouped in pairs, reducing the number of labels on the $x$-axis.}
    \label{fig:stats_of_distance}
\end{figure}

The results show that conditioning directly on endpoint distance is not sufficient to recover the statistical properties of the true intermediates. The same divergence bin can contain trajectories with different elapsed times and different local mutability regimes. As a consequence, distance-conditioned reconstructions are not always correctly calibrated along the evolutionary path. This is especially visible in the distance from the starting sequence, and to a lesser extent in the statistical energy, as shown in Fig.~\ref{fig:stats_of_distance}. In contrast, selecting the time-conditioned model using the inferred timescale gives reconstructions that better match the ground-truth distributions. This supports the conclusion of the main text: endpoint divergence alone is an incomplete proxy for evolutionary time, and topological information from the endpoint sequences is needed to recover the latent dynamical regime.

\section{Inference of latent evolutionary timescales}
\label{sec:timescale_inference}

\subsection{Feature sets for timescale classification}
\label{sec:timescale_feature_sets}

To apply the time-conditioned reconstruction models when the true evolutionary time is unknown, we trained a classifier to infer the timescale $\tau$ from observables computed on the two endpoint sequences. The task is a multiclass classification problem: each trajectory triplet is labeled by the true timescale used in the simulation,
\[
\tau \in \{10^1,10^2,10^3,10^4,10^5,10^6\},
\]
and the classifier predicts this label from features derived from $S_\mathrm{start}$ and $S_\mathrm{end}$.

The datasets for timescale inference were built from the same simulated trajectory triplets described in Section~\ref{subsec:trajectory_triplets} and used in the reconstruction experiments. For each triplet, the midpoint sequence was not used as input to the classifier. Only quantities available from the two endpoints were used, so that the procedure remains applicable to real cases where the intermediate sequence is unknown.

We tested several feature sets. The simplest feature is the Hamming distance between the two endpoints,
\[
d(S_\mathrm{start},S_\mathrm{end}) .
\]
This captures the observed sequence divergence, but it does not contain explicit information about the local topology of the landscape. We then added the average context-dependent entropy of the two endpoints, $\overline{\mathrm{CDE}}_\mathrm{start}$ and $\overline{\mathrm{CDE}}_\mathrm{end}$, which provides a measure of their mutational robustness or local mutability.

We also tested more detailed feature representations. One is the binary mismatch vector between endpoints, where each position is assigned 0 if the two endpoint residues match and 1 if they differ. This preserves the positions of the substitutions, rather than only their total number. Finally, we tested residue-level CDE vectors for the two endpoints, which retain the site-wise mutability profile instead of using only the sequence-averaged CDE values.

Overall, adding CDE information improved timescale inference compared with using endpoint divergence alone. The best performance was obtained by combining the binary mismatch vector with the residue-level CDE vectors of the two endpoints. However, the gain over the simpler representation based on endpoint divergence and average CDE values was modest. For this reason, the main analysis uses the simpler feature set whenever possible, while the full feature comparison is reported here.

\begin{figure}[h]
    \centering
    \includegraphics[width=1\linewidth]{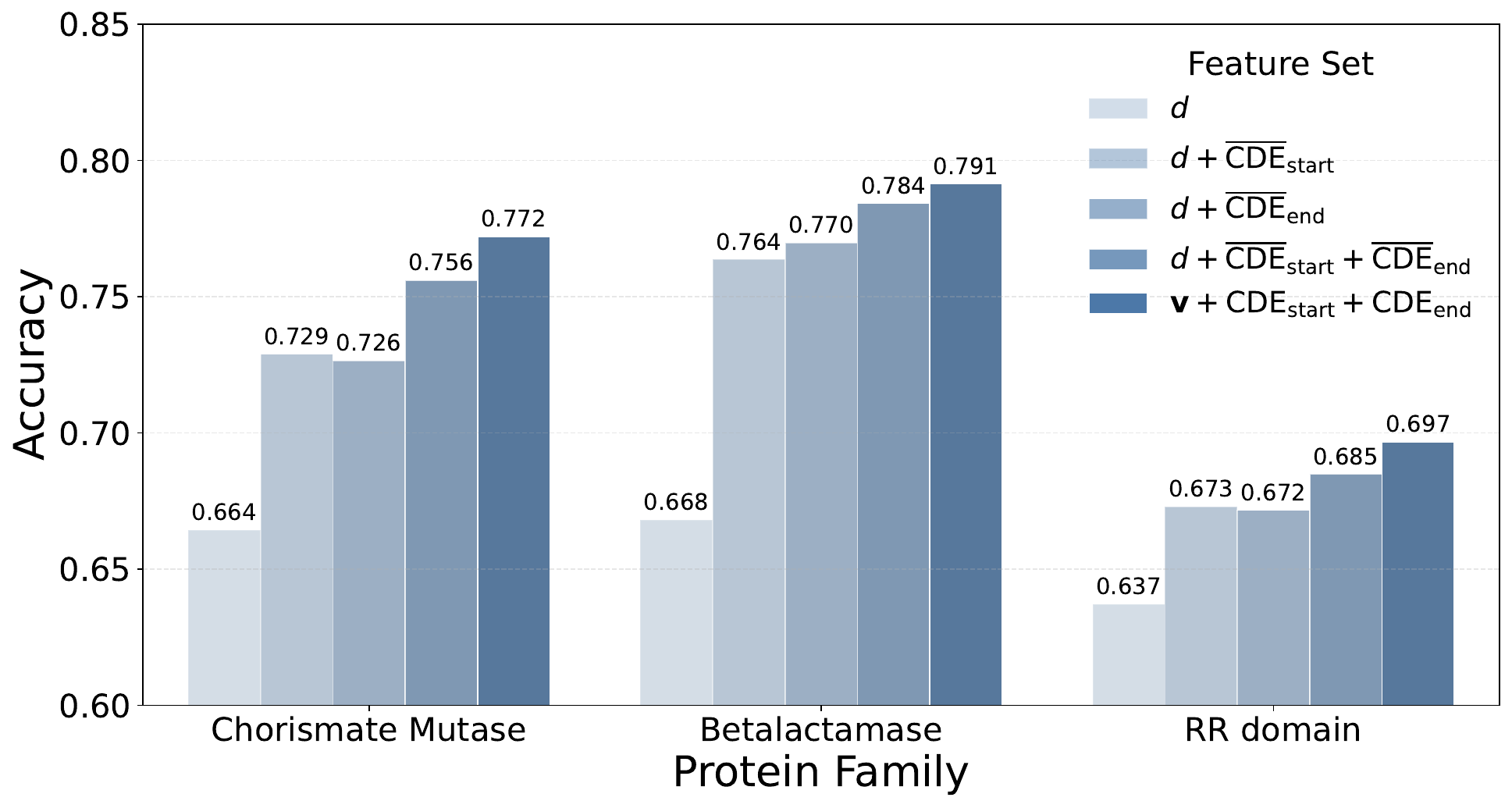}
    \caption{\textbf{Timescale inference from endpoint features across protein families.}
    Performance of multinomial logistic regression classifiers trained to predict the evolutionary timescale $\tau$ from different feature sets. The tested features include endpoint divergence, average $\overline{\mathrm{CDE}}$ values of the start and end sequences, the binary mismatch vector between endpoints, and residue-level CDE profiles. Adding CDE-based information improves timescale inference compared with divergence alone. The full residue-level representation gives only a modest additional gain over simpler CDE-based features.}
    \label{fig:mlr_feature_comparison}
\end{figure}

\end{document}